 \documentclass[final,1p,times]{elsarticle}

\usepackage{lineno}
\usepackage{amssymb, amsfonts}
\usepackage{lipsum}
\usepackage{rotating}
\usepackage{pslatex}
\usepackage{epsfig,soul}
\usepackage[normalem]{ulem}
\usepackage{caption,subcaption}
\usepackage{booktabs,multirow}
\usepackage{graphicx,tikz,graphics,epsfig}
\usepackage{url,CJKutf8}
\usepackage{color,xcolor}
\usepackage{booktabs}
\usepackage{multirow}
\usepackage{setspace} 
\usepackage{physics,amsmath} 
\usepackage{bm}  
\journal{International Journal of Heat and Mass Transfer}

\begin{document}  
\begin{frontmatter}  
%%%% Title, authors and addresses 
%%%% use the tnoteref command within \title for footnotes;
%%%% use the tnotetext command for theassociated footnote;
%%%% use the fnref command within \author or \affiliation for footnotes;
%%%% use the fntext command for theassociated footnote;
%%%% use the corref command within \author for corresponding author footnotes;
%%%% use the cortext command for theassociated footnote;
%%%% use the ead command for the email address,
%%%% and the form \ead[url] for the home page: 
 
\title{Minimum-dissipation model and symmetry-preserving discretization for scalar transport in a turbulent flow} 
 \author[jsun]{Jing Sun\corref{cor1}}
 \ead{j.sun@rug.nl} 
 \author[xavi]{F. Xavier Trias}
 \ead{f.trias@cttc.upc.edu}  
 \author[jsun]{Roel Verstappen}
 \ead{r.w.c.p.verstappen@rug.nl}  
 \cortext[cor1]{Corresponding author}
 \affiliation[jsun]{organization={University of Groningen, Computational and Numerical Mathematics — Bernoulli Institute}, 
            addressline={Nijenborgh 9, 9747 AG, Groningen}, 
            city={Groningen}, 
            country={The Netherlands}}    
 \affiliation[xavi]{organization={Technical University of Catalonia, Heat and Mass Transfer Technological Center}, 
            addressline={C/ Colom 11 Building TR4, Terrassa, 08222}, 
            city={Barcelona}, 
            country={Spain}}    
            
\begin{abstract}    
%%%%\begin{document}   
This work extends the minimum-dissipation model of large-eddy simulation and symmetry-preserving discretization to account for active or passive scalar transport and complex physical mechanisms. This novel scalar-minimum-dissipation model exhibits several desirable properties. It includes the effect of scalar transport, in addition to shear, on the suppression and production of turbulence. It switches off at no-slip walls, ensuring accurate capture of near-wall flow behavior without needing a wall model. Furthermore, it also switches off in laminar and transitional flows so that it is capable of predicting laminar-turbulent transition. 
The new scalar-minimum-dissipation model combined with the symmetry-preserving discretization is successfully tested in a differentially heated cavity in OpenFOAM. The results show that the symmetry-preserving discretization significantly improves predictions of flow quantities and heat transfer on highly stretched meshes. 
\end{abstract}   

\begin{highlights}
\item Novel Scalar-Minimum-Dissipation Model:
Extends the minimum-dissipation model of LES to incorporate active and passive scalar transport, improving turbulence and thermal flow predictions. 
\item Improved Simulations in OpenFOAM:
The symmetry-preserving discretization significantly enhances flow and heat transfer predictions on highly stretched meshes, especially for Rayleigh numbers up to $10^{10}$
\item Efficiency in Coarse Meshes:
At Rayleigh numbers $2\times 10^9$ and $1\times 10^{10}$, the model delivers accurate predictions on coarse meshes, significantly reducing computation time. 
\end{highlights}

\end{frontmatter}

%%\linenumbers
\section{Introduction} 
Large-eddy simulation (LES) is a state-of-the-art numerical approach for studying the transport of momentum and scalars in turbulent flows \cite{Pope_2000}, for instance in applications like geophysical flows \cite{galperin1993large,lesieur2005large} and combustion \cite{pitsch2006large, veynante2002turbulent}. LES resolves turbulent structures larger than a specified filter scale while employing subgrid-scale (SGS) models to represent smaller, unresolved scales \cite{sagaut2005large}. Despite its widespread use, the accuracy of LES in simulating scalar transport flows remains a challenge.  

To enhance LES performance for scalar transport turbulent flow, various SGS models have been proposed. The so-called linear SGS model relates the SGS stress tensor (heat flux) linearly to the large-scale velocity (temperature) gradient, with an eddy viscosity (diffusivity) acting as a proportionality parameter \cite{Eidson1985,schumann1995stochastic}. These linear SGS heat flux models employ a subgrid Prandtl number, $Pr_t$, leading to $\kappa_e=\nu_e/Pr_t$, where $\nu_e$ is the subgrid eddy viscosity. The value of $Pr_t$ is often taken between 0.1 and 1, with 0.6 being the most commonly used \cite{sagaut2005large}. 
However, this constant $Pr_t$ based strategy introduces inherent errors because the subgrid Prandtl number is case-dependent, characterizing the differences that may exist between kinetic energy and scalar spectral transfers \cite{lesieur1989large}. The constant Prandtl number based models can be made more accurate by making the subgrid Prandtl number space and time dependent with the help of dynamic procedures \cite{germano1991,lilly1992proposed,wong1994comparison,Moin1991ADS}. However, these dynamic approaches increase computational complexity and may introduce instability in simulations \cite{Pope_2000,germano1991}. Instability issues can be mitigated by replacing the strain rate tensor with the velocity gradient when applying dynamic procedures to the eddy viscosity model \cite{rozema2022local}. 

An alternative approach to improving linear SGS models involves incorporating additional flow characteristics, such as vorticity and anisotropy, leading to nonlinear models \cite{lund1993parameterization,KOSOVIC1997,PENG20021393,clark1979evaluation}. Among these, the gradient model has garnered significant attention. However, gradient-like models require secondary regularization to ensure numerical stability \cite{bardina1980improved}. Another important consideration in LES is the subgrid closure for source/sink terms when additional physical processes introduce nonlinear dependencies on the scalar field. This issue arises in various applications, such as chemical reactions, temperature-dependent molecular viscosity/diffusivity, and scalar field interactions with microphysics (e.g., radiative transfer) \cite{lund1993parameterization, PENG20021393}.

The minimum-dissipation model has emerged as a promising approach to addressing these challenges. The first minimum-dissipation model (QR), proposed by Verstappen \cite{verstappen2011does, verstappen2008restraining, verstappen2018much}, has been validated in various turbulent flows, including channel flows, periodic hill flows, and flows over cylinders \cite{sun2024}.  Rozema et al. subsequently developed the anisotropic minimum-dissipation model (AMD) for flows on anisotropic grids \cite{rozema2015minimum}. However, previous studies of the QR model have focused primarily on momentum-driven flows. An extension of the AMD model to buoyancy-driven convective boundary layers demonstrated its potential but was limited to specific flow scenarios and relied on mixing-length theory for eddy diffusivity \cite{abkar2017large}.

In this study, we generalize the minimum-dissipation model to turbulent scalar transport flows, accommodating diverse scalar fields such as buoyancy, temperature, and species concentration. The proposed model also addresses the effects of coupled terms in the scalar transport equation. It ensures compatibility with the turbulent viscosity hypothesis without introducing additional assumptions.

To ensure numerical accuracy, the flow system is discretized using a symmetry-preserving finite-volume discretization \cite{arakawa1997computational,vasilyev2000high, manteuffel1986numerical,  perot2011discrete,sanderse2013energy,coppola2023}. Symmetry-preserving discretization, as proposed by Verstappen and Veldman \cite{verstappen2003symmetry}, is designed to exactly preserve the symmetry properties of the underlying differential operators: the discrete convective operator is represented by a skew-symmetric matrix, and the discrete diffusive operator by a symmetric, positive-definite matrix.
Building upon this concept, Verstappen \cite{verstappen2006symmetry} developed and tested a fourth-order symmetry-preserving discretization method for the numerical simulation of turbulent flow and heat transfer in a channel flow with surface-mounted cubes, where temperature is treated as a passive scalar. Trias et al. \cite{TRIAS2014246} generalized Verstappen's method to unstructured collocated meshes, proposing a way to mitigate checkerboard spurious modes, which they tested in differentially heated cavity flows. Komen et al. \cite{komen2021symmetry, janneshopmanRKSymFoam} extended these ideas further by developing a conservative, symmetry-preserving, second-order time-accurate PISO-based pressure-velocity coupling method for solving the incompressible Navier-Stokes equations on unstructured collocated grids in OpenFOAM. Building on these developments, we extend symmetry-preserving discretization to turbulent flows involving active/passive scalar transport. 

The new minimum-dissipation model and  symmetry-preserving discretization are implemented and verified in OpenFOAM using simulations of a differentially heated cavity, where scalar gradients are coupled with momentum via buoyancy terms in the momentum equation. This test case is highly relevant to engineering applications such as room ventilation, electronic cooling, and building airflow. Experimental and numerical studies, such as those by Trias et al. \cite{trias2007}, provide valuable benchmarks for testing numerical schemes and turbulence models.

In summary, this work addresses the limitations of LES in turbulent scalar transport by generalizing the minimum-dissipation model and symmetry-preserving discretization. The novel approach improves LES accuracy for scalar transport turbulent flows, bridging a key gap in existing modeling techniques.

This paper is organized as follows: Sections \ref{sec:MDM} and \ref{sec:symmetry} generalize the QR model and symmetry-preserving discretization schemes to turbulent flows involving passive/active scalars, respectively. Section \ref{sec:case} presents and discusses a buoyancy-driven turbulent flow example. Finally, Section \ref{sec:sum} summarizes the results and conclusions. 

\section{Scalar transport minimum-dissipation model} \label{sec:MDM}
\subsection{Governing equations}
The dynamics of incompressible Newtonian fluid with constant physical properties in turbulent flow, involving an active or passive scalar, or an additional source/sink term, can be described by the continuity, momentum, and scalar transport equations, written as
\begin{alignat}{2}
     \label{eq:basicEqn}     
	 \nonumber\nabla \cdot u &= 0, \\ 
     \partial_t  u + ( u\cdot \nabla)  u &= &&\nu \nabla^2u + \underbrace{f(u,\theta)}_{source/coupling\hspace{0.3em} term}-\nabla  p,  \\ \nonumber
     \partial_t  \theta + ( u\cdot \nabla)  \theta &= &&\kappa\nabla^2\theta + \underbrace{g(u,\theta)}_{source/coupling \hspace{0.3em}term}, 
\end{alignat}
where $u$, $p$, and $\theta$ represent the velocity, kinematic pressure, and scalar quantity (e.g., temperature, pollutant concentration, salinity), respectively. The parameters $\nu$ and $\kappa$ denote the kinematic molecular viscosity and molecular diffusivity, respectively. The generic forms $f(u,\theta)$ and $g(u,\theta)$ describe the coupling.

LES seeks to predict the dynamics of spatially filtered turbulent flows. By applying a spatial filter with filter length $\delta$ to the governing equations \eqref{eq:basicEqn}, we obtain  
\begin{alignat}{3}
     \partial_t \bar u + &\nabla\cdot( \bar u \bar u^T) && =\nu \nabla^2\bar u -\nabla \cdot \left(\overline{u u^T}- \bar u\bar u^T\right) &&+ \overline{f(u,\theta)} -\nabla \bar p,\quad \nabla \cdot \bar u = 0,	\nonumber\\
     \partial_t \bar \theta + &\nabla\cdot(\bar u\bar \theta) && = \kappa \nabla^2\bar \theta - \nabla\cdot(\overline{u\theta}-\bar u\bar\theta) &&+ \overline{g(u,\theta)}.
     \label{eq:filterLES}
\end{alignat}
Here, $\bar{u}$ and $\bar{\theta}$ denote the spatially filtered velocity and scalar fields, respectively. The filter operations $u \mapsto \bar{u}$ and $\theta \mapsto \bar{\theta}$ are assumed to commute with differentiation. $\bar{p}$ represents the filtered pressure and $(\,)^T$ refers to the transpose of a matrix.   
The subgrid stress tensor $\overline{u'u'}$ and subgrid scalar flux $\overline{u' \theta'} $ are defined as
\begin{alignat}{3}
\overline{u' u'} &= \overline{uu^T} -\bar u \bar u^T,\quad&&u'= u-\bar u, \nonumber\\
\overline{u' \theta'} &= \overline{u\theta} -\bar u \bar \theta,  \quad&&\theta'= \theta-\bar \theta.
\end{alignat}
The velocity-scalar covariance $\overline{u'\theta'}$ is a vector, representing the flux of the scalar due to the residual of the velocity field. The scalar fluxes in the filtered scalar equation play an analogous role to that of  subgrid stress tensor. In particular, they give rise to a closure problem: even if $\bar u$ is known, the filtered scalar equation cannot be solved for $\bar \theta$ without a  model for $\overline{u'\theta'}$. 
To close the filtered equations \eqref{eq:filterLES}, closure models $\tau(v)$ and $\tau_\theta(v)$ are introduced, leading to
\begin{alignat}{2}
     &\partial_t v + \nabla\cdot(vv^T) &&=\nu \nabla^2\bar u-\nabla \cdot \tau(v) + \tilde f(v,\vartheta;\delta) - \nabla \tilde p , \quad  \nabla \cdot v = 0 \nonumber\\
     &\partial_t \vartheta + \nabla\cdot(v\vartheta) &&=\kappa \nabla^2 \vartheta  - \nabla \cdot \tau_\theta(v) + \tilde g(v,\vartheta;\delta), 
     \label{eq:les}
\end{alignat}
where the variable name is changed from $\bar u$ to $v$ and $\bar\theta$ to $\vartheta$ to emphasize that solutions of \eqref{eq:les} differ from those of \eqref{eq:filterLES}. $\tilde f$ and $\tilde g$ are closures of $f$ and $g$ with $\overline{f(u,\theta)}\approx\tilde f(\bar u, \bar \theta; \delta)\approx \tilde f(v,\vartheta;\delta)$ and $\overline{g(u,\theta)}\approx\tilde g(\bar u, \bar \theta; \delta)\approx \tilde g(v,\vartheta;\delta)$, respectively. $\tilde p=\bar p + \frac{1}{3}kI$ is the modified pressure, where the residual kinetic energy $k$ is defined to be half of the trace of the subgrid stress tensor $k\equiv\frac{1}{2}\langle \overline{u'u'}\rangle$ and $I$ is the identity matrix; $\tau(v)\equiv \overline{u'u'} -  \frac{1}{3}kI$ is the approximation of the anisotropic part of the subgrid stress tensor and $\tau_\theta(v)$ is the approximation of the subgrid scalar flux.

\subsection{Subgrid-scale model}
Using the turbulent-viscosity and gradient-diffusivity hypotheses, $\tau(v)$ and $\tau_\theta(v)$ can be expressed as:
\begin{align}
    \label{eq:tauij}
   &\tau(v)= -2\nu_e S(v), \nonumber \\
 &\tau_\theta(v) = -\kappa_{e}\nabla \vartheta,
\end{align}
where $S(v) = (\nabla v + \nabla v^T)/2$ is the symmetric part of the velocity gradient, with zero trace due to $\nabla \cdot v = 0$. The coefficients $\nu_e$ and $\kappa_e$ are the eddy viscosity and eddy diffusivity, respectively. This subgrid model is time-irreversible (for $\nu_e > 0$ and $\kappa_e > 0$), forward in time it provides dissipation and diffusion, i.e., the complexity of the flow can be reduced, depending on the eddy viscosity and diffusivity.
In common practice, $\kappa_e$ is related to $\nu_e$ through the subgrid Prandtl number $Pr_t$, such that $\kappa_e=\nu _e /Pr_t$.
This relationship is used in the test cases presented in this paper. Note that this approach differs from Ref. \cite{abkar2017large}, where the mixing-length approximation yields $\kappa_e =C_S^2\delta^2S/ Pr_t$. Nevertheless, using the same concept of the minimum-dissipation model, we derive a minimum-diffusivity model for LES.

Consider a box filter over an arbitrary grid cell $\Omega_\delta$ of the flow domain with diameter $\delta$. The filtering operators $u \mapsto \bar{u}$ and $\theta \mapsto \bar{\theta}$ are defined as
\begin{align}
   & \bar u = \frac{1}{|\Omega _\delta |}\int_{\Omega_\delta} u(x,t) dx, \nonumber\\
    &\bar \theta = \frac{1}{|\Omega _\delta |}\int_{\Omega_\delta} \theta(x,t) dx.
\end{align}
The filter length $\delta$ is the user's chosen length of the filter. In practice, the filter is often related to the computational grid by choosing the filter length $\delta$ equal to the mesh spacing, so that the filter solution can be captured on the computational grid. 
The filtered velocity $\bar{u}$ and scalar $\bar{\theta}$ is equal to the averages of $u$ and $\theta$  over $\Omega_\delta$, respectively.

Minimum-dissipation model assumes that the eddy viscosity/diffusivity model must keep the residual fields $v'=v-\bar v$ and $\vartheta' =  \vartheta -\bar \vartheta$ from becoming dynamically significant. This condition is formalized by confining the subgrid energy with Poincaré's inequality. 
Poincaré's inequality indicates that there exists a constant $C_\delta$, depending only on $\Omega_\delta$, such that for every function $v$ and $ \vartheta $ in the Sobolev space $W^{1,2}(\Omega_\delta)$ 
\begin{align}
\label{eq:poincare}
    &\int_{\Omega_\delta}\frac{1}{2}\parallel v'\parallel^2 dx \leq C_\delta \int_{\Omega_\delta} \frac{1}{2}\parallel \nabla v \parallel ^2 dx, \quad
    v'= v - \frac{1}{\Omega_\delta} \int_{\Omega_{\delta}} v dx\\ \nonumber
    &\int_{\Omega_\delta}\frac{1}{2}\parallel \vartheta'\parallel^2 dx \leq C_\delta \int_{\Omega_\delta} \frac{1}{2}\parallel \nabla \vartheta \parallel ^2 dx, \quad
    \vartheta'= \vartheta -\frac{1}{\Omega_\delta} \int_{\Omega_{\delta}} \vartheta dx
\end{align}
where the residual fields $v'$ and $\vartheta'$ contain the eddies of size smaller than the length of the filter $\delta$, and $\parallel \cdot \parallel = \sqrt {\langle, \rangle}$ is the standard norm of the inner product $\langle, \rangle$ on the space of real valued $L^2(\Omega_\delta)$ functions. 
The Poincaré constant $C_\delta$, independent of $v$ and $\vartheta$, equals the inverse of the smallest non-zero eigenvalue of the dissipative operator $-\nabla^2 = - \nabla \cdot \nabla = \nabla ^T \nabla$ on the grid cell $\Omega_\delta$ \cite{courant2008methods}. Payne and Weinberger have derived a continuous value of the Poincaré constant $C_\delta = \delta^2/\pi^2$ for convex domains of diameter $\delta$ \cite{payne1960optimal}. 

The subgrid energy on the left-hand side of Eq.\eqref{eq:poincare} is unknown in a LES simulation, but the evolution of the velocity gradient energy and scalar gradient energy can be determined by applying the gradient operator to Eq.\eqref{eq:les} and multiplying the resulting equations by $\nabla v$ and $\nabla \vartheta$, respectively. Integration by parts gives 
\begin{align} 
     \label{eq:gradientEnergy}
        \nonumber\frac{d}{dt} \int_{\Omega_\delta} \frac{1}{2} \parallel \nabla v \parallel ^2 dx =& -\nu \int_{\Omega_\delta} \parallel \nabla^2 v\parallel ^2 dx + \int_{\Omega_\delta} (v \cdot \nabla) v \cdot \nabla^2 v dx \\ \nonumber
       & -\nu_e \int_{\Omega_\delta} \parallel \nabla^2 v \parallel^2 dx + \int_{\Omega_\Delta} \tilde f \cdot \nabla^2 v dx,\\
     \frac{d}{dt} \int_{\Omega_\delta} \frac{1}{2} \parallel \nabla \vartheta \parallel ^2 dx =& -\kappa \int_{\Omega_\delta} \parallel \nabla^2 \vartheta \parallel ^2 dx - \int_{\Omega_\delta}  (\nabla v \nabla\vartheta)\cdot \nabla\vartheta dx \\\nonumber
     &- \kappa_e \int_{\Omega_\delta} \parallel \nabla^2 \vartheta \parallel^2 dx  + \int_{\Omega_\Delta} \tilde g \cdot \nabla^2 \vartheta dx \nonumber
\end{align}
where boundary terms that result from the integration by parts vanish because $\Omega_\delta$ is a periodic box. Physically, boundary integrals represent transport of momentum and scalar gradient energy rather than dissipation or production and are therefore neglected in deriving the minimum-dissipation model. 
 
The source or coupling terms $\tilde f$ and $\tilde g$ in Eq.\eqref{eq:gradientEnergy} can be decomposed into components parallel and perpendicular to $\nabla^2 v$ and  $\nabla^2 \vartheta$, respectively
\begin{align}
   \label{eq:sourceDecomp}
   &\tilde f =\frac{\tilde f\cdot \nabla^2  v}{||\nabla^2  v||^2}\nabla^2  v + \tilde f^{\perp}\\ \nonumber
   &\tilde g =\frac{\tilde g\cdot \nabla^2  \vartheta}{||\nabla^2  \vartheta||^2}\nabla^2  \vartheta + \tilde g^{\perp},
\end{align}
where $\tilde f^{\perp}$ and $\tilde g^\perp$ are perpendicular to the velocity Laplacian $\nabla^2 v$  and scalar Laplacian $\nabla^2\vartheta$, respectively, and therefore do not contribute to changes in gradient energy.

The principle behind the minimum-dissipation and minimum-diffusivity models is that the production of gradient energy is balanced by the subgrid model so that the gradient energy dissipates at a natural rate, i.e. by molecular dynamics and the external forces.  
The second terms on the right-hand side of Eq.\eqref{eq:gradientEnergy} represent gradient energy production by convection and the third terms represent dissipation and diffusion due to eddy viscosity and diffusivity, respectively. 
Now suppose the eddy viscosity and eddy diffusivity in Eq.\eqref{eq:gradientEnergy} are taken such that the subgrid-scale models cancel out the production of gradient energy from the convective term and source/coupling terms, then we obtain 
\begin{align}
    &\nu_e \int_{\Omega_\delta} \parallel \nabla^2 v \parallel^2 dx =\int_{\Omega_\delta} (v \cdot \nabla) v \cdot \nabla^2 v dx + \int_{\Omega_\delta} \frac{\tilde f\cdot \nabla^2  v}{||\nabla^2  v||^2}\nabla^2  v \cdot \nabla^2 v dx, \\ \nonumber
   &\kappa_e \int_{\Omega_\delta} \parallel \nabla^2 \vartheta \parallel^2 dx = \int_{\Omega_\delta} -(\nabla v\cdot \nabla\vartheta) \nabla\vartheta dx + \int_{\Omega_\delta} \frac{\tilde g\cdot \nabla^2  \vartheta}{||\nabla^2  \vartheta||^2}\nabla^2  \vartheta \cdot \nabla^2 \vartheta  dx.
\end{align}
By doing this, the scales smaller than the grid size stop producing. Giving the minimum-dissipation and -diffusion conditions
\begin{align}
    &\frac{d}{dt} \int_{\Omega_\delta} \frac{1}{2} \parallel \nabla v \parallel ^2 dx = -\nu \int_{\Omega_\delta} \parallel \nabla^2 v\parallel ^2 dx, \\
    &\frac{d}{dt} \int_{\Omega_\delta} \frac{1}{2} \parallel \nabla \vartheta \parallel ^2 dx = -\kappa \int_{\Omega_\delta} \parallel \nabla^2 \vartheta\parallel ^2 dx. 
     \label{eq:7}
\end{align}
The production of velocity gradient energy by the convective term can be rewritten as in Ref. \cite{verstappen2011does}
\begin{equation}
(v\cdot \nabla) v\cdot \nabla^2 v = 4r(v) + \nabla \cdot (...),
\end{equation}
where 
\begin{equation}
r(v) = -det(S) = -\frac{1}{3} tr(S^3) = -\frac{1}{3}SS:S
\end{equation}
is the third invariant of the resolved strain-rate-tensor. The subgrid dissipation rate of velocity gradient energy at the scale of the filter box $\Omega_\delta$ can be approximated by assuming that the eddy viscosity is constant over the filter box. Then the application of the Poincaré inequality yields
\begin{equation}
\nu_e\int_{\Omega_{\delta}} \lVert\nabla^2 v\rVert^2 dx= \nu_e\int_{\Omega_{\delta}}q(\omega)dx \geq \frac{ 4 \nu_e}{C_\delta}\int_{\Omega_{\delta}}q(v)dx ,
\end{equation}
where the $\omega=\nabla \times v$ and
\begin{equation}
q(v)= \frac{1}{2}tr(S^2) = \frac{1}{2}S:S
\end{equation}
is the second invariant of the resolved strain-rate-tensor. For the detailed derivation of $r(v)$ and $q(v)$, see \cite{verstappen2011does,verstappen2018much,rozema2015minimum}. 

Applying Poincaré inequality to the scalar Laplacian term, we obtain
\begin{equation}
    \int_{\Omega_\delta} \parallel \nabla \vartheta \parallel^2 dx \leq  C_\delta \int_{\Omega_\delta} \parallel \nabla^2 \vartheta \parallel^2 dx.
\end{equation}
Thus, eddy-viscosity (diffusivity) models give sufficient eddy dissipation (diffusion) to cancel the production of the velocity (scalar) gradient energy, respectively if the inequalities
\begin{alignat}{2}
&  \int_{\Omega_\delta} \frac{\tilde f\cdot \nabla^2  v}{\lVert\nabla^2  v\rVert^2}\nabla^2  v \cdot \nabla^2 v dx +4\int_{\Omega_\delta}r(v)dx  \leq 4 \frac{\nu_e}{C_{\delta}}\int_{\Omega_\delta}q(v)dx\\ \nonumber
& \int_{\Omega_\delta} \frac{\tilde g\cdot \nabla^2  \vartheta}{\lVert\nabla^2  \vartheta\rVert^2}\nabla^2  \vartheta \cdot \nabla^2 \vartheta  dx+ \int_{\Omega_\delta} -(\nabla v\cdot \nabla\vartheta) \nabla\vartheta dx \leq \frac{ \kappa_e}{C_\delta} \int_{\Omega_\delta} \parallel \nabla \vartheta \parallel^2 dx  
\end{alignat} 
hold. 

The minimum eddy dissipation/diffusion that satisfies these conditions is given by 
\begin{align}
&\nu_e=C_\delta\frac{\int_{\Omega_\delta}r(v)dx +\frac{1}{4} \int_{\Omega_\delta} \frac{\tilde f\cdot \nabla^2  v}{\lVert\nabla^2  v\rVert^2}\nabla^2  v \cdot \nabla^2 v dx }{\int_{\Omega_\delta} q(v) dx}\\
&\kappa_e=C_\delta\frac{\int_{\Omega_\delta} -(\nabla v\cdot \nabla\vartheta) \nabla\vartheta dx + \int_{\Omega_\delta} \frac{\tilde g\cdot \nabla^2  \vartheta}{\lVert\nabla^2  \vartheta\rVert^2}\nabla^2  \vartheta \cdot \nabla^2 \vartheta  dx}{\int_{\Omega_\delta} \parallel \nabla \vartheta \parallel^2 dx}
\end{align} 
The midpoint rule is used to approximate the integrals and a numerical approximation of the Poincaré constant gives the eddy viscosity and eddy diffusivity of the scalar-QR model 
\begin{align}
	\label{eq:scalarQR}
&\nu_e=C\delta^2\frac{max\big\{ r(v) +\frac{1}{4} \frac{\tilde f\cdot \nabla^2  v}{\lVert\nabla^2  v\rVert^2}\nabla^2  v \cdot \nabla^2 v, \hspace{0.5em}0\big\}}{ q(v) }\\ \nonumber
&\kappa_e = C \delta^2 \frac{max\big\{-(\nabla v \nabla\vartheta)\cdot \nabla\vartheta + \frac{\tilde g\cdot \nabla^2  \vartheta}{\lVert\nabla^2  \vartheta\rVert^2}\nabla^2  \vartheta \cdot \nabla^2 \vartheta,0\big\}} {\parallel \nabla \vartheta \parallel^2},
\end{align}
where $C$ is the model coefficient and $\delta$ is the LES filter width. In practice, the negative values of the numerator are clipped off since the small scales of motion are stopped and cannot produce large scales of motion. 
 
On anisotropic grids, the anisotropic minimum-dissipation model is proposed to compensate for the error introduced by the anisotropic grid \cite{rozema2015minimum}. In the AMD model, the model coefficient does not depend on the grid size, instead, the grid size dependence is incorporated into the gradient operator, yielding the scalar-AMD model
\begin{align}
	\label{eq:scalarAMD}
&\nu_e= C_a\frac{max\big\{(-\Lambda \cdot \nabla v)^T(\Lambda \cdot \nabla v):S +  \frac{\tilde f\cdot \nabla^2  v}{\lVert\nabla^2  v\rVert^2}\nabla^2  v \cdot (\Lambda\cdot\nabla)^2 v, 0\big\}}{ \nabla v : \nabla v} \\ \nonumber
&\kappa_e = C_a\frac{max\big\{(-\Lambda\cdot\nabla v )^T(\Lambda\cdot\nabla\vartheta)\cdot \nabla\vartheta +  \frac{\tilde g\cdot \nabla^2  \vartheta}{\lVert\nabla^2  \vartheta\rVert^2}\nabla^2  \vartheta \cdot (\Lambda\cdot \nabla)^2 \vartheta,0\big\}} {\parallel \nabla \vartheta \parallel^2},
\end{align}
where $\Lambda$ represents a diagonal matrix with its diagonal elements containing the cell size $\delta_i$ (i=1,2,3) in each direction. Note that the coupling terms $\tilde f$ and $\tilde g$ in the scalar-QR \eqref{eq:scalarQR} and scalar-AMD models \eqref{eq:scalarAMD} have no contribution when $\tilde f \cdot \nabla^2 v$ is non-positive, as the non-positive term does not generate velocity gradient energy. However, when $\tilde f \cdot \nabla^2 v > 0$, the coupling terms contribute to the production of gradient energy. To ensure that motions at scales smaller than the filter size stop producing energy, this term must be counteracted by subgrid-scale models. Additionally, the scalar-QR and scalar-AMD models are applicable to both passive and active scalar transport. In the case of a passive scalar, no feedback terms exist in the governing equations from the scalar transport, resulting in $\tilde f$ and $\tilde g$ vanishing. When the scalar is active, for instance, salinity and temperature scalars can interact to contribute to energy changes, requiring the model to adjust accordingly. 

\subsection{Physical importance}
The novel scalar-QR \eqref{eq:scalarQR} and scalar-AMD \eqref{eq:scalarAMD} LES models exhibit several desirable properties. They incorporate the effects of scalar transport, in addition to shear, on the suppression and production of turbulence. They switch off at no-slip walls, ensuring accurate capture of near-wall flow behavior without the need for wall models. Additionally, they switch off in laminar flows, enabling accurate prediction of laminar-turbulent transitions. Unlike traditional models, these approaches do not require ad hoc stability or shear corrections \cite{abkar2017large}. The model coefficient $C$ remains consistent across both the SGS viscosity and diffusivity models, eliminating the need to determine separate optimal coefficients for the scalar transport equation.
\section{Numerical method}\label{sec:symmetry}
\subsection{Symmetry-preserving spatial discretization} 
The numerical schemes used to discretize the governing equations as well as the eddy viscosity \eqref{eq:scalarQR} and eddy diffusivity \eqref{eq:scalarAMD} do maintain symmetry properties of the underlying differential operators. To achieve this, we extended symmetry-preserving discretization \cite{verstappen2006symmetry, TRIAS2014246,komen2021symmetry, HOPMAN2025113537, santos2025} in OpenFOAM for the scalar transport in turbulent flow.
These discretizations preserve the underlying symmetry properties of the continuous differential operators. Specifically, the convective operator is represented by a skew-symmetric matrix, and the diffusive operator by a symmetric positive semi-definite matrix. Such discrete operator properties guarantee stability and ensure that the global kinetic energy balance is exactly preserved, even on coarse meshes, provided the incompressibility condition is satisfied and time-step is small enough to guarantee the stability of the time integration.
As a result, the kinetic energy is not systematically damped by the discrete convective term, nor does it need to be damped explicitly to ensure the stability of the method. This is a crucial point because artificial dissipation would interfere with the subtle balance between convective transport and physical dissipation, which usually affects the smallest scales of motion, the essence of turbulence (modeling).

\subsubsection{Symmetry-preserving discretization of the Navier--Stokes equation}
Following the same notation as in \cite{TRIAS2014246}, the time evolution of the discrete velocity vector $\mathbf{u_c}$ is governed by the finite-volume discretization of the incompressible Navier-Stokes equations 
\begin{align}  
    \bm{\Omega}\frac{\mathrm d\mathbf{u_c}}{\mathrm dt} +\mathbf{C(u_s)u_c + Du_c} + \bm{\Omega}\mathbf{G_c p_c} + \bm{\Omega} \mathbf{f_c=0_c}, \quad \quad \mathbf{Mu_s = 0_c}
    \label{eq:discreteNS}
\end{align}
where subscripts $s$ and $c$ denote whether the variables are staggered at the cell faces or cell-centered. The term $\mathbf{f_c} \in \mathbb{R}^{3n}$ includes external forces, such as feedback terms and the Coriolis force.  The $\mathbf{p_c} \in \mathbb{R}^{n}$ and $\mathbf{u_c} \in \mathbb{R}^{3n}$ are the cell-centered pressure and velocity fields, and $n$ is the number of control volumes. The auxiliary discrete staggered velocity field $\mathbf{u_s} \in \mathbb{R}^{m}$ is related to the cell-centered velocity through the linear transformation $\Gamma_{c \rightarrow s} \in \mathbb{R}^{m \times 3n}$,  such that $\mathbf{u_s} \equiv \Gamma_{c \rightarrow s} \mathbf{u_c}$, where $m$ is the number of faces in the computational domain. 
The matrix $\bm{\Omega} \in \mathbb{R}^{3n \times 3n}$ is a positive-definite diagonal matrix representing the sizes of the control volumes, the convective coefficient matrix $\mathbf{C}(\mathbf{u_s}) \in \mathbb{R}^{3n \times 3n}$ is skew-symmetric, and the discrete diffusive operator $\mathbf{D} \in \mathbb{R}^{3n \times 3n}$ is a symmetric positive-definite matrix. $\mathbf{M} \in \mathbb{R}^{n \times m}$ is the face-to-center discrete divergence operator, and $\mathbf{G_c} \in \mathbb{R}^{3n \times n}$ is the discrete gradient operator. For the detailed construction of the discrete operators, see Ref. \cite{TRIAS2014246}.

\subsubsection{Symmetry-preserving discretization of the scalar transport equation}
\paragraph{Basic discrete operators}
In matrix-vector notation, the time evolution of the discrete scalar field $\bm{\theta_c}$ is governed by the finite-volume discretization of the scalar transport equation \eqref{eq:basicEqn} 
\begin{equation} 
   \bm{\Omega_c}\frac{\mathrm d\bm{\theta_c}}{\mathrm {d}t} + \mathbf{C_c(u_s)}\bm{\theta_c} + \mathbf{ D_c}\bm{\theta_c} + \mathbf{g_c =0_c},
   \label{eq:discreteScalar} 
\end{equation}
where $\bm{\theta_c} \in \mathbb{R}^{n}$ is the cell-centered scalar variable, and $\bm{\Omega_c} \in \mathbb{R}^{n \times n}$ is the positive-definite diagonal matrix representing the sizes of the cell-centered control volumes. The terms $\mathbf{C_c(u_s)} \in \mathbb{R}^{n \times n}$ and $\mathbf{D_c} \in \mathbb{R}^{n \times n}$ denote the cell-centered convective and diffusive operators for the discrete scalar field, respectively, while $\mathbf{g_c} \in \mathbb{R}^n$ accounts for the source or sink terms in the scalar transport equation. 

\paragraph{Symmetry and conservation properties}
We define the discrete inner product as follows 
\begin{equation}
\langle\bm{\phi_c}, \bm{\theta_c} \rangle =\bm{\phi_c}^* \bm{\Omega_c \theta_c}.
\end{equation} 
Then, the global discrete scalar energy (the $\mathrm{L2}$ norm of the scalar field $\bm{\theta_c}$) is defined as $\lVert \bm{\theta_c}\rVert^2\equiv\bm{\theta}\mathbf{^*_c}\bm{\Omega_c\theta_c}$. Its temporal evolution  can be derived by left-multiplying Eq.\eqref{eq:discreteScalar} by  $\bm{\theta_c}^*$ and summing the resulting expression with its conjugate transpose 
\begin{equation}
\frac{\mathrm d}{\mathrm dt}\lVert \bm{\theta_c}\rVert^2= -\bm{\theta}\mathbf{^*_c\bigl(C_c(u_s)+C^*_c(u_s)\bigr)}\bm{\theta_c} -\bm{\theta}\mathbf{^*_c(D_c +D^*_c)}\bm{\theta_c}.
\label{eq:discreteL2Scalar}
\end{equation}
In the absence of diffusion (i.e., $\mathbf{D_c} = 0$), the global scalar energy is conserved if and only if the convective terms vanish in the discrete scalar energy equation
\begin{equation}
\bm{\theta}\mathbf{^*_c\bigl(C_c(u_s)+C^*_c(u_s)\bigr)}\bm{\theta_c} =0,
\label{eq:skewScalar}
\end{equation}
of any solution $\bm{\theta_c}$ of the dynamic system \eqref{eq:discreteScalar}. This equality holds if and only if the discrete convective operator is skew-symmetric 
\begin{equation}
\vb{C_c(u_s)}=-\vb{C_c^*(u_s)}.
\label{eq:skewConv}
\end{equation} 
Moreover, in this way, the skew-symmetry of the convective continuous operator is preserved. So, in conclusion, if the convective operator is properly chosen, the global scalar energy reduces to 
\begin{equation}
\frac{\mathrm d}{\mathrm dt}\lVert \bm{\theta_c}\rVert^2= -\bm{\theta}\mathbf{^*_c(D_c +D^*_c)}\bm{\theta_c} \leq0,  
\end{equation}
where the inequality arises from the condition that the diffusive term must be strictly dissipative. Thus, the matrix $\mathbf{D_c + D^*_c}$ must be positive-definite. Although not strictly necessary for dissipation of energy, the diffusive operator $\mathbf{D_c}$ is also assumed to be symmetric, consistent with its continuous counterpart $-\Delta = -\nabla \cdot \nabla$. 

In conclusion, the scalar energy $\lVert \bm{\theta_c} \rVert^2$ of the discrete system \eqref{eq:discreteScalar} does not increase over time if the discrete convective operator $\mathbf{C_c(u_s)}$ is skew-symmetric and $\mathbf{D_c + D^*_c}$ is positive semi-definite. Under these conditions, the semi-discrete system \eqref{eq:discreteScalar}  is stable, and a solution to \eqref{eq:discreteScalar} can be obtained on any computational grid without the need for artificial damping mechanisms to stabilize the spatial discretization. 

\subsubsection{Constructing the discrete operators}
\begin{figure}[b!]
\centering
\includegraphics[trim=355 180 355 205, clip, width=0.35\linewidth]{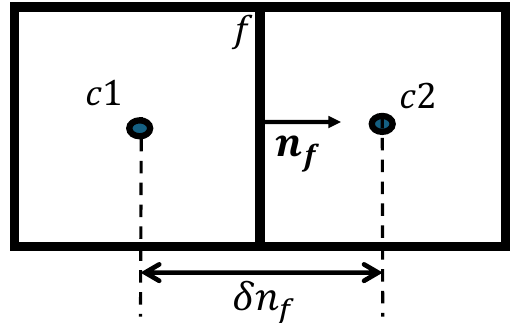}
\captionsetup{font={footnotesize}}
\caption{Control volume for structured collocated grids.}
\label{fig:collocatedGrid}
\end{figure} 
On a collocated grid, both the scalar and velocity variables are stored at the center of the control volume, whereas a secondary scalar and velocity are defined at the faces to represent fluxes (see Figure \ref{fig:collocatedGrid}). 
\paragraph{Conservation of mass}
Since mass and scalar quantities are transported at the same velocity, the mass flux is employed to discretize the convection of the scalar $\bm{\theta_c}$. The finite-volume method begins with the integral form of the mass conservation equation. Integrating the continuity equation in \eqref{eq:basicEqn} over an arbitrary centered cell $k$ of volume $(\bm{\Omega_c})_{k,k}$ yields 
\begin{equation}
    \int\limits_{(\Omega_c)_{k,k}} \nabla \cdot \vb{u} \,\mathrm dV=\int\limits_{\partial(\Omega_c)_{k,k}}\vb{u}\cdot \vb{n}\,\mathrm dS=\sum_{f\in F_f(k)}\int\limits_{S_f}\vb{u}\cdot \vb{n}\,\mathrm dS,
    \label{eq:fvmContinuity}
\end{equation} 
where the summation is over the faces $f$ of the cell $k$, together constituting the set $F_f(k)$, $\mathbf{n}$ is the outward-pointing normal. The discretization of the integral in \eqref{eq:fvmContinuity} is given by 
\begin{equation}
    \sum_{f\in F_f(k)}\int\limits_{S_f}\vb{u}\cdot \vb{n}\,\mathrm dS \approx \sum_{f\in F_f(k)} [\vb{u}_s]_f A_f,
    \label{eq:approxDivU}
\end{equation}
where $A_f$ is the area of the face $f$ and $[\mathbf{u_s}]_f$ approximates the normal velocity at the faces $f$.  Using this, conservation of mass becomes
\begin{equation}
    [\vb{M}\vb{u}_s]_k \equiv  \sum_{f\in F_f(k)} [\vb{u}_s]_f A_f =0,
    \label{eq:discreteContinuity}
\end{equation}
where $\vb M$ is the integrated divergence operator. 
\paragraph{Skew-symmetry of the convective operator}
In finite-volume discretization, the convective term of the scalar transport in Eq. \eqref{eq:basicEqn} can be expressed as 
\[
  \int\limits_{(\Omega_c)_{k,k}} \nabla \cdot (\mathbf{u}\bm{\theta}) \, \mathrm dV=\int\limits_{\partial(\Omega_c)_{k,k}}\bm{\theta}\vb{u}\cdot \vb{n}\,\mathrm dS= \sum_{f\in F_f(k)}\int\limits_{S_f}\bm{\theta}\vb{u}\cdot \vb{n} \,\mathrm dS \approx\sum_{f\in F_f(k)}[\bm{\theta_c}]_f[\vb{u_s}]_f A_f,
    \label{eq:fvmConvective}
 \]  
where $\bm{[\theta_c]}_f$ denotes the approximation of the scalar flux at face $f$. The convective operator is thus defined by its action on an arbitrary cell-centered scalar field $\bm{\theta_c} \in \mathbb{R}^{n}$ at some cell $k$ as
\begin{equation}
[\mathbf{C_c(u_s)}\bm{\theta_c}]_k=\sum_{f\in F_f(k)}\bm{[\theta_c]}_f[\vb{u_s}]_f A_f 
\end{equation}
For this to hold, it is necessary that the $[\bm{\theta_c}]_f$ fluxes are approximated by an equal-weighted interpolation between the faces of the control volume: 
\begin{equation}
[\bm{\theta_c}]_f = \frac{1}{2}\bigl([\bm {\theta_c}]_{c1} +[\bm {\theta_c}]_{c2}\bigr),
\label{eq:midPoint}
\end{equation}
where $c1$ and $c2$ denote the control volumes adjacent to the face $f$ (see Figure \ref{fig:collocatedGrid}). Recall that the discrete normal velocity $\mathbf{[u_s]}_f$ is also  computed via equal-weighted interpolation
\begin{equation}
    [\mathbf{u_s}]_f = \frac{1}{2}\bigl([\mathbf{u_c}]_{c1} +[\mathbf{u_c}]_{c2}\bigr)\cdot \vb{n}_f,
\end{equation}
in order to arrive at Eq.\eqref{eq:skewConv}.\\
On nonuniform grids, one might attempt to tune the weight $\frac{1}{2}$ in the interpolation \eqref{eq:midPoint} to the local mesh size in order to minimize the order of the local truncation error. In doing so, the skew symmetry of the convective differential operator is lost. Therefore, we take constant weights of $\frac{1}{2}$, even on non-uniform meshes. 

Furthermore,  for skew-symmetry of $\mathbf{C_c(u_s)}$, the diagonal elements must be zero, i.e., 
\begin{equation}
    [\mathbf{C_c(u_s)}]_{k,k}=\frac{1}{2}\sum_{f\in F_f(i)}[\vb{u}_s]_f A_f=0,
\end{equation}
which holds because the  discrete divergence of $\mathbf{u_s}$ vanishes, see Eq.\eqref{eq:discreteContinuity}. 
Hence, the collocated convective operator at cell $k$ is given as
\begin{equation}
    \mathbf{[C_c(u_s)\bm{\theta_c}]}_k = \sum_{f \in F_f(k)}\frac{1}{2}\bigl([\bm{\theta_c}]_{c1} +[\bm{\theta_c}]_{c2}\bigr)[\mathbf{u_s}]_f A_f.
\end{equation} 

\paragraph{Diffusive operator}
The diffusive operator is formulated as the product of two first-order differential operators: the divergence and the gradient. To ensure symmetry and positive definiteness of the diffusive operator, $\mathbf{D_c}$, the divergence is discretized in a finite volume fashion as described in \eqref{eq:fvmContinuity}–\eqref{eq:discreteContinuity}. 
The integrated gradient operator, $\bm{\Omega_s }\mathbf{G_c}$, is represented by the negative transpose of the discrete divergence, $\mathbf{-M^*}$ \cite{TRIAS2014246}. Then, the coefficient matrix $\mathbf{D_c}$ is given by 
\begin{equation}
    \mathbf{D_c} =-\frac{1}{\mathrm{RePr}}\vb{MG_c}=\frac{1}{\mathrm{RePr}}\vb{M}\bm{\Omega_s}^{-1}\mathbf{M}^*,
\end{equation} 
where $\mathrm{Re}$ and $\mathrm{Pr}$ are the Reynolds and Prandtl numbers, respectively. This construction leads to a symmetric, positive-definite approximation of the diffusive operator $-\nabla \cdot \nabla$. The collocated diffusive operator on a cell-centered variable $\bm{\theta_c}$ is thus given by
\begin{equation}
    [\mathbf{D_c}\bm{\theta_c}] =\frac{1}{\mathrm{RePr}}\sum_{f\in F_f(k)}\frac{(\bm{\theta}_{c1}-\bm{\theta}_{c2})A_f}{\delta n_f},
\end{equation}
where the length $\delta n_f$ is an approximation of the distance between the centroids of cells $c1$ and $c2$, given by $\delta n_f = |\mathbf{n}_f \cdot \overrightarrow{c1c2}|$ (see Figure \ref{fig:collocatedGrid}). This is essentially the same $\mathbf D$ as in the discrete momentum equations \eqref{eq:discreteNS}, but of course applied to a scalar not to a vector. 

\subsection{Time integration}
The governing equations for turbulent flow and heat transfer are integrated in time using the second-order Crank-Nicolson method. Further details on the Crank-Nicolson method can be found in \cite{komen2021symmetry}. On a collocated grid the decoupling of the velocity and pressure introduces an additional error to the momentum equation. Trias et al. \cite{TRIAS2014246} proposed a solution for this decoupling problem without introducing any non-molecular dissipation. The idea behind this approach is to use a linear shift operator to transform a cell-centered velocity $\vb{u_c}$ into a staggered one $\mathbf{u_s}$ and use an incremental-pressure projection method for the explicit time stepping.  Subsequently, Komen et al. \cite{komen2021symmetry} developed a symmetry-preserving, second-order, time-accurate projection and PISO-based methods to achieve nearly conservative results on collocated grids. This paper adopts the methodology developed by Komen et al.  \cite{komen2021symmetry}. The pressure correction method for both implicit and explicit time stepping methods, are thoroughly detailed in \cite{komen2021symmetry}. 
 
\subsection{OpenFOAM solver} 
Numerical schemes are implemented in OpenFOAM and the solver is called buoyantBoussinesqRKSymFoam. 
The buoyantBoussinesqRKSymFoam solver’s primary algorithm comprises three hierarchical iterative levels in the case of implicit time discretization \cite{komen2021symmetry,janneshopmanRKSymFoam}:\\
1. The outer loop iterates over each Runge--Kutta stage.  (Here the Butcher array is taken such that the Runge--Kutta method coincides with Crank-Nicolson.) \\
2. The intermediate loop manages to update the nonlinear convective term.\\
3. The inner PISO loop handles the pressure-velocity coupling.\\ 

\section{Test-case: differentially heated cavity} \label{sec:case}
Turbulent thermal stratification plays an important role in the atmospheric and oceanic flow. Thermally stratified fluid forms layers by temperature. Temperature difference and gravity are the driving mechanism. The temperature difference causes the colder and heavier fluid to settle at the bottom while allowing the warmer and lighter fluid to float. Thermal stratification can strongly affect the mixing of fluids, which is important in many manufacturing processes.  Therefore, we take thermally stratified fluid as an example to assess the proposed symmetry-preserving discretization and scalar-minimum-dissipation model.  Only the scalar-QR model is assessed in this work; the assessment of the scalar-AMD is encouraged for interested readers. 
\begin{figure}[!b]
    \centering
    \includegraphics[trim=275 80 370 50, clip, width=0.31\linewidth]{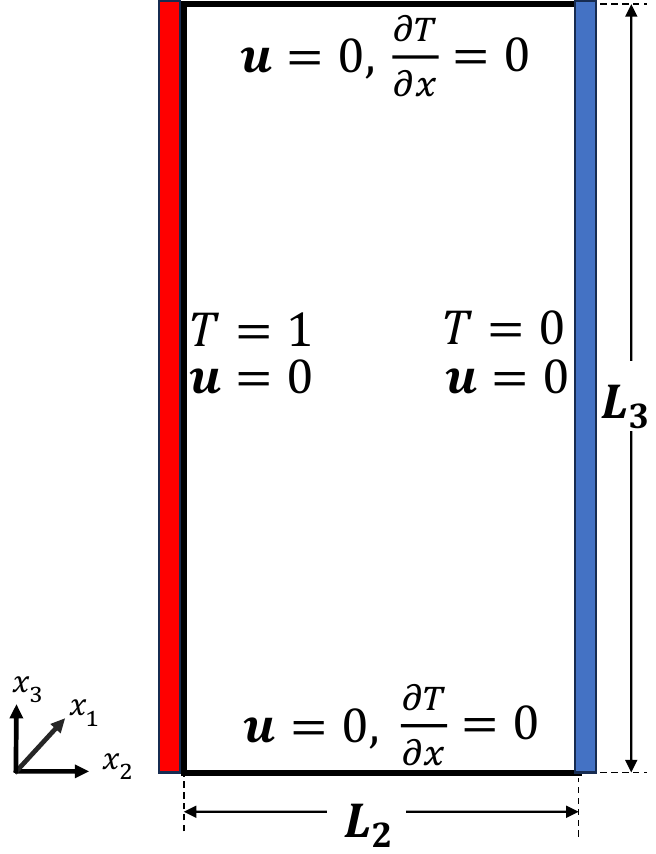}
    \caption{Problem set-up for the differentially heated cavity.}
    \label{fig:geometry}
\end{figure} 

The coordinate system is defined as follows (Figure \ref{fig:geometry}): $x_1$ represents the periodic direction, while $x_2$ and $x_3$ are allocated to denote the horizontal and vertical directions, respectively, corresponding to the two wall-normal dimensions. The height and depth aspect ratios are expressed as $L_3/L_2=4$ and $L_1/L_2=1$. Note the Boussinesq approximation is used to account for the buoyancy effects  and the thermal radiation is neglected.\\
The cavity is subjected to a temperature difference across the vertical isothermal walls $(T (x_1$, $ 0$, $ x_3) = 1$, $T (x_1$, $1$, $x_3) = 0)$ while the top and bottom walls are adiabatic. No-slip boundary conditions for velocity are enforced at four specific planes: $x_2 = 0$, $x_2 = 1$, $x_3 = 0$, and $x_3 = L_3/L_2$. Periodic boundary conditions are utilized in the $x_1$-direction.\\
In accordance with prior work conducted by Soria et al. \cite{soria2004} and Trias et al. \cite{trias2007}, the simulation assumes that the cavity is filled with fluid, characterized by a laminar Prandtl number (Pr) of 0.71, and has a height aspect ratio ($L_3/L_2$) equal to 4. The configuration is solely dependent on the Rayleigh number (Ra) and the depth aspect ratio ($L_1/L_2$).
\subsection{Governing equations}
We consider an incompressible viscous flow with constant thermal conductivity and viscosity, with Coriolis force, and with gravity acting in the $L_3$-direction (i.e., the gravitational acceleration is given by $\textbf{g}=(0,0,-g)$ (see Figure \ref{fig:geometry}). Note that the Coriolis force is typically not considered in differentially heated cavity simulations. However, it may be of interest in other applications, such as geophysical flows. The conservation of mass and energy are filtered as in Eq.\eqref{eq:les}. The active scalar $\vartheta$ represents temperature and $\tau_\theta(v)$ refers to the subgrid heat flux. The feedback term by $\vartheta$ is buoyancy. Using the Boussinesq assumption (i.e., the density $\rho$ is assumed to vary only with temperature), we obtain the momentum equation of LES
\begin{equation}
\partial_t v + (v\cdot \nabla) v =- \nabla \tilde p + 2\nu \nabla \cdot S(v)  -\nabla \cdot \tau(v) + \underbrace{[1-\beta(\vartheta-\vartheta_0)]\mathbf{g}}_{feedback\hspace{0.3em} term} + f_c\epsilon v
\label{eq:temp}
\end{equation} 
where $\vartheta_0$ is the reference temperature; $f_c$ is the Coriolis parameter; $\epsilon$ is the alternating unit tensor; $\mathbf{g}$ is the gravitational acceleration; $\beta$ is the thermal expansion coefficient
\begin{equation}
\beta \equiv -\frac{1}{\rho}\frac{\partial \rho}{\partial \vartheta}\approx -\frac{1}{\rho_0}\frac{\rho-\rho_0}{\vartheta-\vartheta_0},
\end{equation}
where $\rho_0$ is the reference density at the reference temperature. Thermal radiation is neglected. Note that the closure is not needed for feedback terms since the linearity in $\vartheta$ and $v$. Substituting the source term $\tilde f$ in Eq.\eqref{eq:scalarQR} with the feedback term and the Coriolis force in Eq.\eqref{eq:temp} yields 
\begin{align}  
    \nu_e &= C \delta^2 \frac{max\big\{r(v)+\frac{1}{4} g\beta\nabla v_g \cdot \nabla(\vartheta -\vartheta_0),0\big\}}{q(v)} \nonumber \\
    \label{eq:tempVisc}
    \kappa_e &= C\delta^2\frac{max\{-(\nabla v \nabla\vartheta)\cdot \nabla\vartheta, \,0\}} {\parallel \nabla \vartheta \parallel^2},
\end{align}
where $v_g$ is the velocity component in the gravitational acceleration direction. The Coriolis force vanishes, see \ref{appendixA}. The scalar-QR eddy-viscosity model includes stratification effects, the effect of buoyancy, in addition to shear, on the suppression and production of turbulence. 

\subsection{Physical and numerical parameters}
\subsubsection{Dimensionless parameters}
The Rayleigh number based on the cavity height is given by
\begin{equation}
    \mathrm{Ra}=\frac{g\beta d\vartheta L_3^3}{\nu \kappa},
\end{equation}
where $d\vartheta$ is the temperature difference. The Nusselt number $Nu$ is the dimensionless form of the convection heat transfer coefficient and provides a measure of the convection heat transfer at a solid surface. The reference heat flux is given by $\lambda d\vartheta /L_3$, where $\lambda$ is the thermal conductivity. Thus, the dimensionless local Nusselt number at the vertical hot wall is given by $Nu(x_3)=-\partial \vartheta /\partial x_2|_{x_2=0}$. 
The hot wall-averaged Nusselt number is given by $Nu=\int_{0}^{1}Nu(x_3)dx_3$ \cite{trias2007}. To describe the structure of the cavity core, the dimensionless stratification $C_{str}$ is defined as $C_{str}=\partial \vartheta /\partial x_3 |_{x_2=0.5L_2/L_3,x_3=0.5}$. The Brunt-Väisälä frequency or (buoyancy frequency) is used as a measure of the stability of a fluid to vertical displacements. It is defined as $N=(C_{str}Pr)^{0.5}/2\pi$. The wall shear stress at the hot wall is defined as $Ra^{-1/4}\frac{\partial  u_3}{\partial x_2}|_{x_2=0}$.

\subsubsection{Numerical settings}
\begin{table}[b!]
    \centering
    \footnotesize
    \begin{tabular}{cccccccccc}
    \hline
         Case&  Ra& $N_1$&    $N_2$&    $N_3$&  $(\Delta x_2)_{min}$&    Total Time & Average Time \\
         \hline
%%         A1-5WM& $6.4\times 10^8$& 64& 144&    318&    $7.58\times 10^{-4}$&   $4.5\times 10^{-3}$&    500&    350 \\
%%         A6WM& $6.4\times 10^8$& 64& 144&    318&    $7.58\times 10^{-4}$&   $4.5\times 10^{-3}$&    500&    330 \\
%%         A6SW& $6.4\times 10^8$&  64& 144&    318&    $7.58\times 10^{-4}$&   $4.5\times 10^{-3}$&    500&    300 \\
%%         A6NWM& $6.4\times 10^8$&  64& 144&    318&    $7.58\times 10^{-4}$&   $4.5\times 10^{-3}$&    420&    220 \\
%%         A7WM(S)& $6.4\times 10^8$&  64& 144&    318&  $ 4.82\times 10^{-5}$&   $2.5\times 10^{-3}$&   240& 40 \\
%%         A7WM& $6.4\times 10^8$&  64& 144&    318&     $ 4.82\times 10^{-5}$&   $2.5\times 10^{-3}$&    500&    300 \\
%%         A7Chorin& $6.4\times 10^8$& 64& 144&    318&     $ 4.82\times 10^{-5}$&   $3.0\times 10^{-3}$&    500&    300 \\
%%         A7vanKan& $6.4\times 10^8$& 64& 144&    318&     $ 4.82\times 10^{-5}$&   \color{blue}$1.4\times 10^{-2}$&    500&    300 \\
%%         A8NWM& $6.4\times 10^8$&  64& \color{blue}194&    \color{blue}460&    $2.78\times 10^{-5}$&   $2.2\times 10^{-3}$&    500&    300 \\
%%         A9NWMC& $6.4\times 10^8$& 64& 156& 312& $2.06\times 10^{-4}$&   $1.26\times 10^{-3}$&    500&    300 \\ 
         A10& $6.4\times 10^8$& 64& 156& 312&  1.63e-3& 4030&    1800 \\ 
         A11& $6.4\times 10^8$& 128& 156& 312& 1.63e-3& 3450&    1200 \\ 
         B1& $2.0\times 10^9$&  64& 144& 318& 1.27e-3& 3350&    2010  \\  
         C1& $1.0\times 10^{10}$& 128& 138&   326&9.61e-4& 550&   350 \\
%         C2& $1.0\times 10^{10}$&  128& 138&   326&9.61e-4& 2500&  1000\\
%         C3& $1.0\times 10^{10}$&  128& 138&   326&9.61e-4& 2500&  1000\\
%%         E1& $1.0\times 10^{11}$&  128& 384&   682&9.4714e-4& $--e-3$&    5000&    4000 \\
%%         E2& $1.0\times 10^{11}$&  64& 196&   384&9.47e-4&  250&    50 \\
         DNS$^1$& $6.4\times 10^8$& 128& 156&    312&2.44e-4&  1000&    800 \\
         DNS$^2$& $2.0\times 10^9$& 64& 144&    318&1.88e-4&   800&    550 \\
         DNS$^3$& $1.0\times 10^{10}$&   64& 138&   326&   1.36e-4&  800&    550 \\
%%         DNS$^5$& $1.0\times 10^{11}$&   128& 682&   1278&  1.36e-4&    800&    550 \\
         DNS$^4$& $ 1.0\times 10^{10}$&  128& 190&    462&    9.63e-5& 440&  300\\ 
         \hline 
    \end{tabular}
    \captionsetup{font={footnotesize}}
	    \caption{Physical and numerical parameters. The variables ($N_1, N_2, N_3$) represents the number of grid points.  "Total time" indicates the cumulative integration period of the Navier-Stokes equations from initial conditions. "Average time" is the time spent on collecting the statistics. These times are non-dimensionalized using the reference time $t_{ref}=L_3/u_{ref}$.  DNS$^1$, DNS$^2$, and DNS$^3$ correspond to cases A, B, and C in Table 1 of Ref. \cite{trias2007}, respectively. DNS$^4$ refers to case C in Table 1 of Ref. \cite{TRIAS2010674}.}
	    %	   DNS$^4$ is the case E in Table 1 of Ref \cite{TRIAS2010665}.
    \label{tab:dhc1}
\end{table}
The second-order symmetry-preserving is employed for spatial discretization. The temporal discretization is the second-order Crank-Nicolson scheme. The time-step is controlled by limiting the maximum CFL number to 0.4. Statistical data on the flow properties are collected after a sufficiently long integration period to achieve a statistically steady-state behavior.  The pressure solver is the GAMG solver from OpenFOAM with DIC smoother. The velocity and temperature are solved by the PBiCGStab solver with a DILU preconditioner. The fractional-step method is used to solve the velocity-pressure coupling with two predictor and four corrector iterations. The grid spacing is constant in the periodic direction, while the grid in the wall-normal (both vertical and horizontal) directions is distributed using a fixed stretching ratio $SR=\frac{\Delta x_{max}}{\Delta x_{min}}$. Averages over the two statistically invariant coordinates (time and $x_1$-direction) are carried out for all fields. 

\subsection{Results and discussion} 
Differential heated fluid-filled cavities at Rayleigh number (based on the cavity height) $\mathrm{Ra}=6.4\times 10^6$, $2\times 10^9$ and $1\times 10^{10}$ are considered. The physical and numerical parameters for all these cases are documented in Table \ref{tab:dhc1}. Table 1 also provides the reference  parameters established by Trias et al. \cite{trias2007,TRIAS2010665,TRIAS2010674} for comparison.  
\subsubsection{Validation of the numerical method}
\begin{figure}[!b]
    \centering
    \includegraphics[trim=142 15 142 13, clip, width=.158\linewidth]{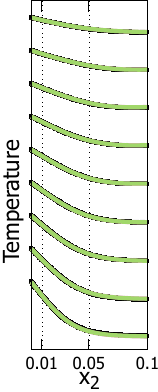}
    \includegraphics[trim=145 15 140 13, clip, width=.156\linewidth]{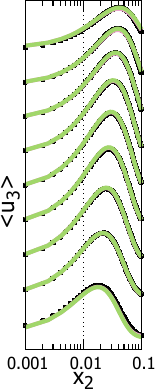}
    \includegraphics[trim=140 15 142 13, clip, width=.1615\linewidth]{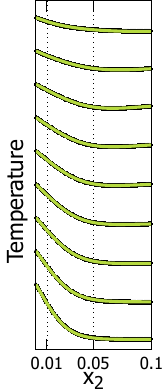}
    \includegraphics[trim=140 15 142 13, clip, width=.1605\linewidth]{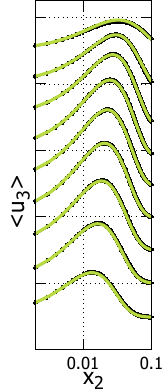}
    \includegraphics[trim=140 16 142 13, clip, width=.161\linewidth]{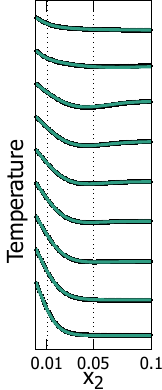}
    \includegraphics[trim=140 5 136 0, clip, width=.156\linewidth]{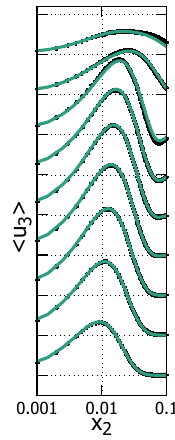}
    \captionsetup{font={footnotesize}}
    \caption{The averaged temperature (left-hand side of the pair) and vertical velocity (right-hand side of the pair) profiles at $x_3=0.4$, $0.8$, $1.2$, $1.6$, $2.0$, $2.4$, $2.8$, $3.2$, and $3.6$ obtained from the simulations with $\mathrm{Ra}=6.4\times 10^8$ (left), $\mathrm{Ra}=2.0\times 10^9$ (middle) and $\mathrm{Ra}=1\times 10^{10}$ (right). Each vertical subdivision represents 0.5 units.}
    \label{fig:ABC}
\end{figure}
 \begin{figure}[!b]
    \centering
    \includegraphics[trim=68 0 72 0, clip, width=.3\linewidth]{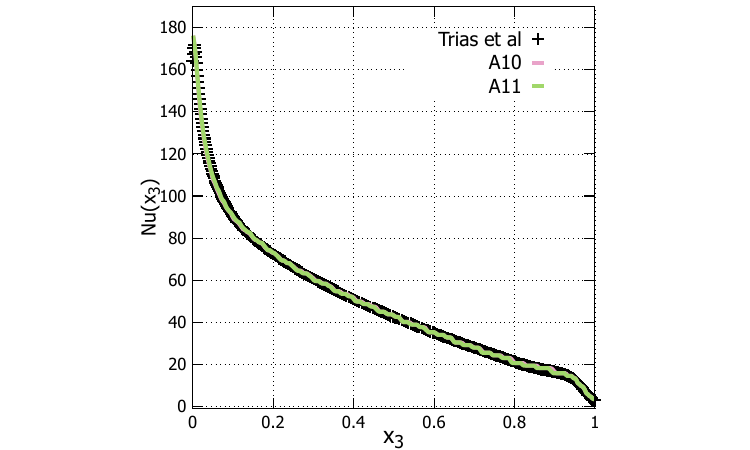}
    \includegraphics[trim=62 8 72 0, clip, width=.32\linewidth]{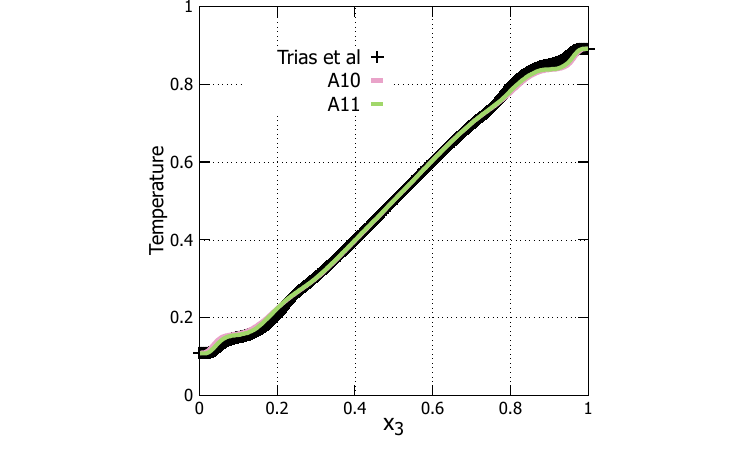}
    \includegraphics[trim=60 0 72 0, clip, width=.31\linewidth]{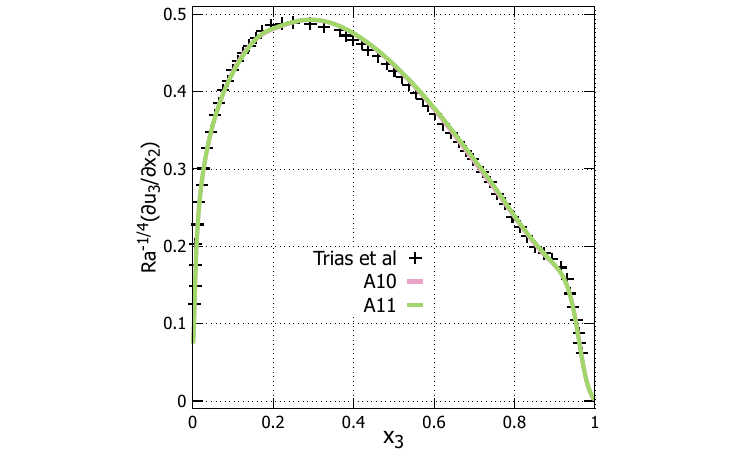}
    \includegraphics[trim=78 15 78 0, clip, width=.32\linewidth]{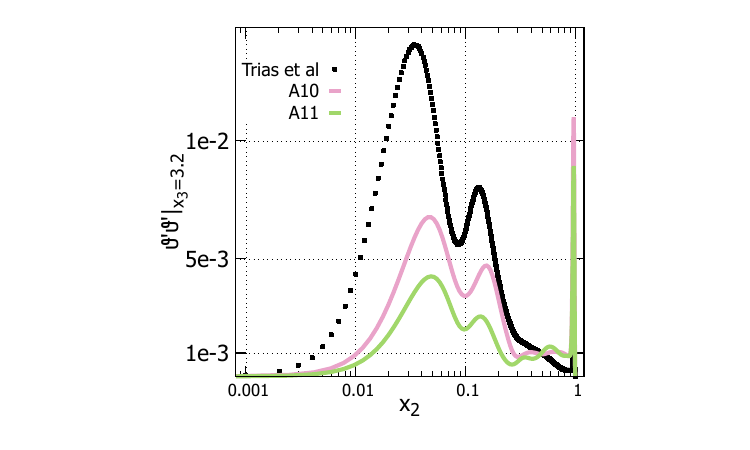}
    \includegraphics[trim=78 15 75 10, clip, width=.325\linewidth]{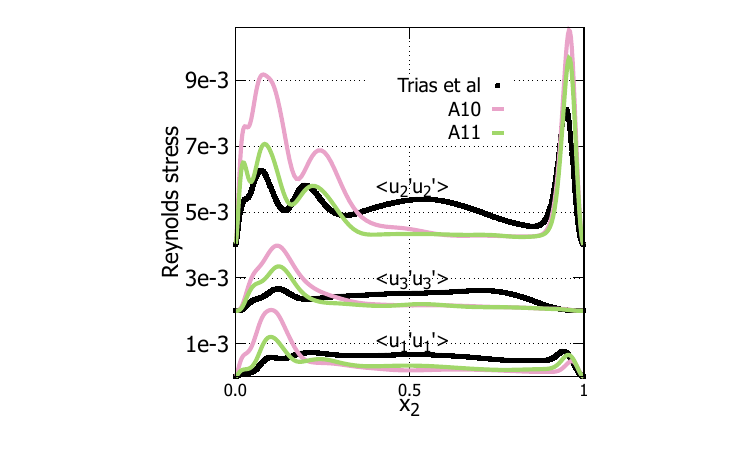}
    \includegraphics[trim=67 5 75 0, clip, width=.31\linewidth]{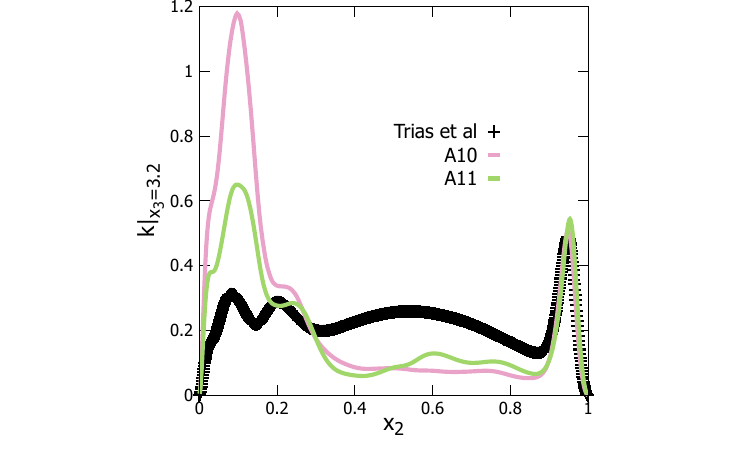}
    \captionsetup{font={footnotesize}}
    \caption{Averaged flow variables and heat transfer  at $\mathrm{Ra}=6.4\times 10^8$. The local Nusselt number distribution at the hot wall $Nu(x_3)|_{x_2=0}$; The averaged temperature vertical profile at mid-width cavity ($x_2=0.5,x_1=0.5$); The averaged wall-shear stress scaled by $Ra^{-1/4}$ at the hot wall, $\frac{\partial u_3}{\partial x_2}|_{x_2=0}$; Horizontal profile at $x_3=3.2$ of the temperature fluctuation $\vartheta'\vartheta'$, velocity fluctuations ($u'_1u'_1$, $u'_2u'_2$ and $u'_3u'_3$) and the turbulence kinetic energy $k$.}
    \label{fig:Ra8_1}
\end{figure}
\begin{figure}[!t]
    \centering
    \includegraphics[trim=68 0 72 0, clip, width=.3\linewidth]{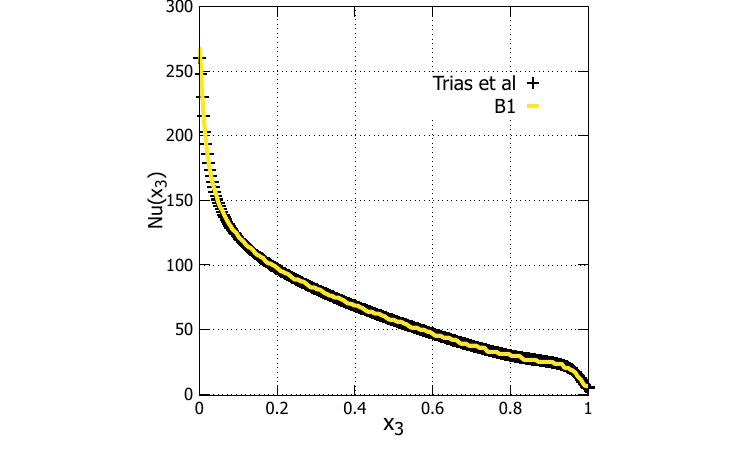}
    \includegraphics[trim=68 5 72 0, clip, width=.31\linewidth]{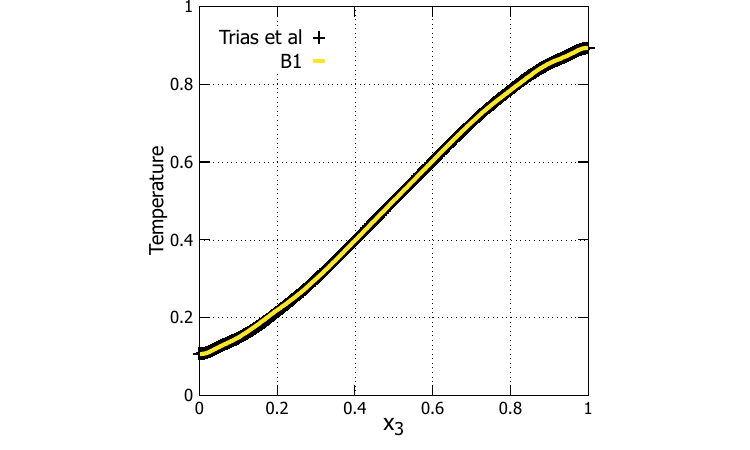}
    \includegraphics[trim=62 0 72 2, clip, width=.31\linewidth]{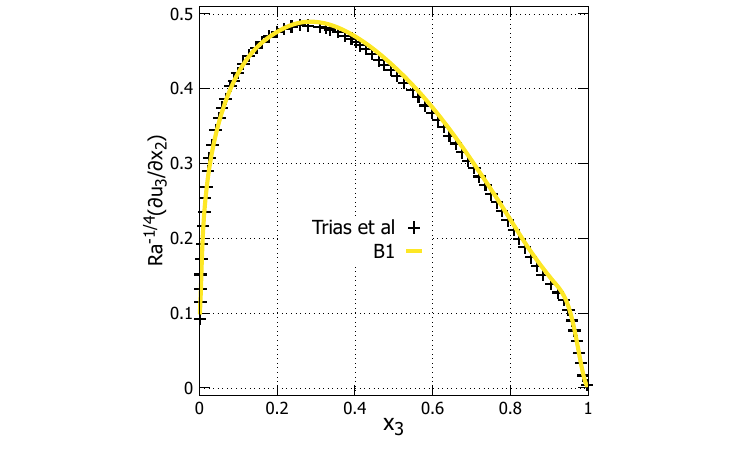}
    \includegraphics[trim=58 0 75 0, clip, width=.31\linewidth]{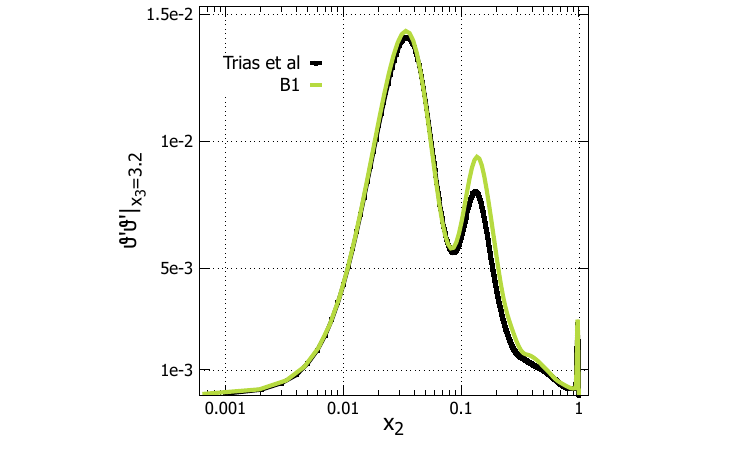}
    \includegraphics[trim=58 0 75 0, clip, width=.31\linewidth]{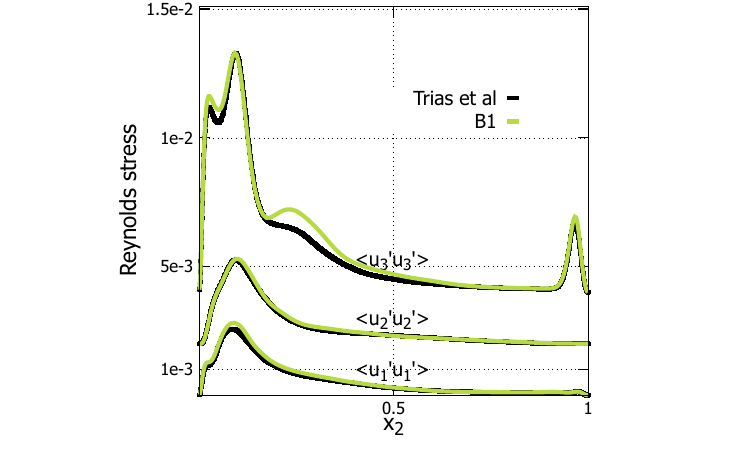}
    \includegraphics[trim=60 0 75 0, clip, width=.31\linewidth]{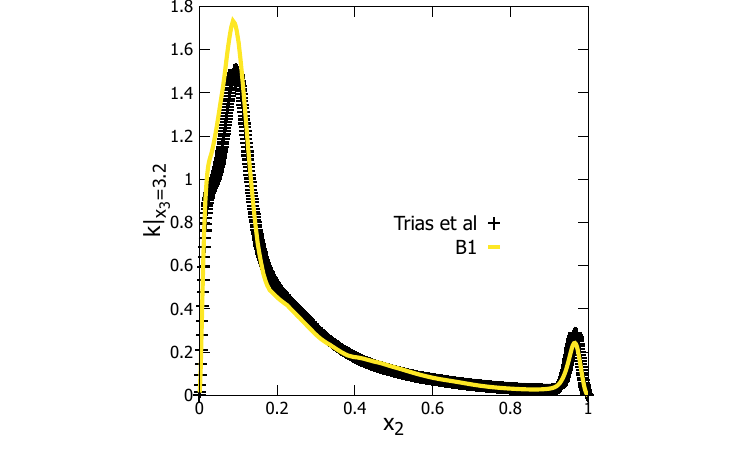}
    \captionsetup{font={footnotesize}}
    \caption{Averaged flow variables and heat transfer at $\mathrm{Ra}=2.0\times 10^9$. }
    \label{fig:Ra9_B2_2} 
    \centering
    \includegraphics[trim=68 0 72 0, clip, width=.31\linewidth]{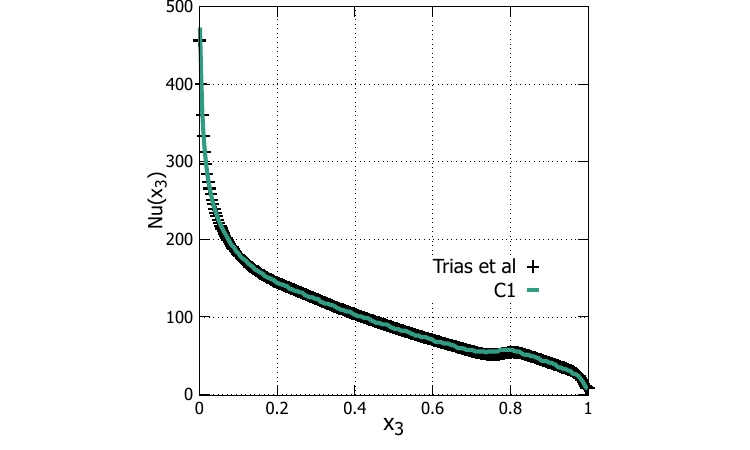}
    \includegraphics[trim=62 0 72 0, clip, width=.31\linewidth]{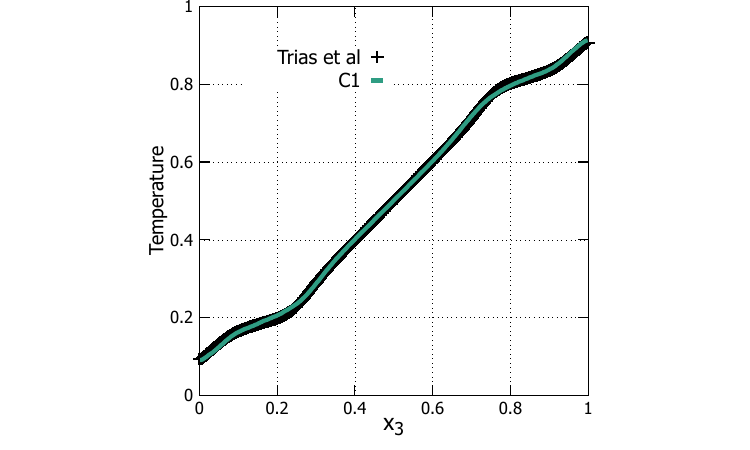}
    \includegraphics[trim=60 0 75 0, clip, width=.31\linewidth]{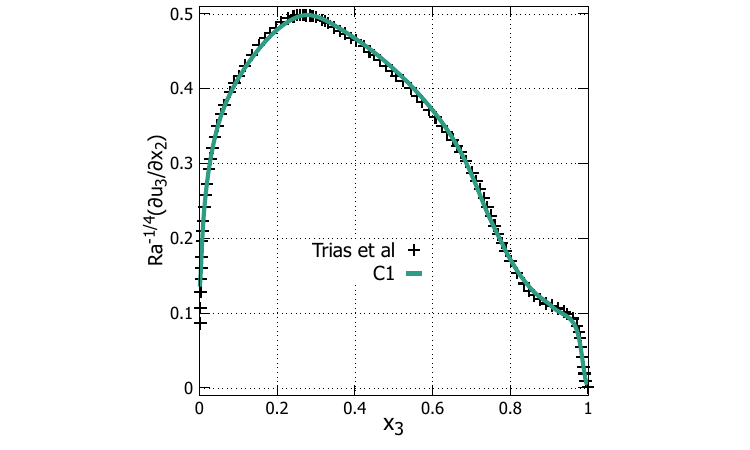}
    \includegraphics[trim=68 12 75 8, clip, width=.315\linewidth]{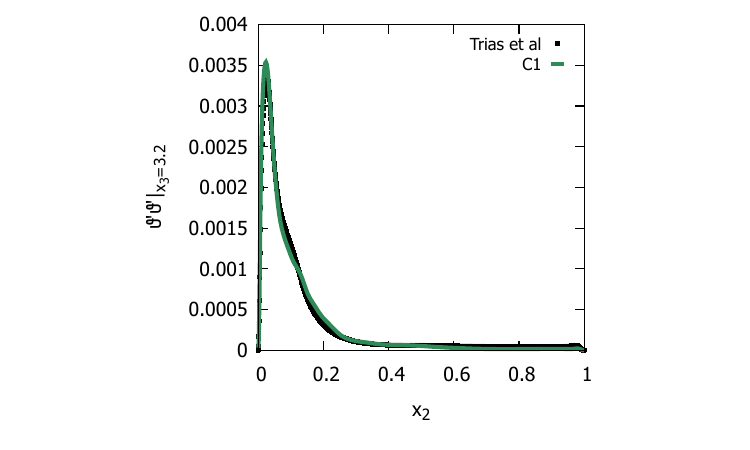}
    \includegraphics[trim=72 10 72 12, clip, width=.31\linewidth]{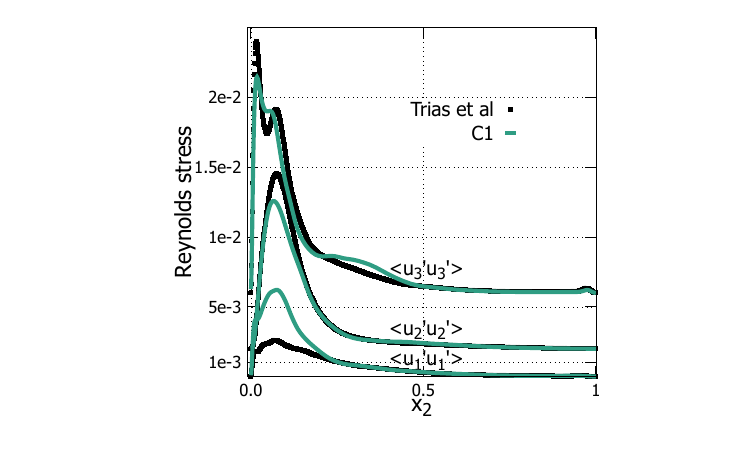}
    \includegraphics[trim=70 10 82 8, clip, width=.315\linewidth]{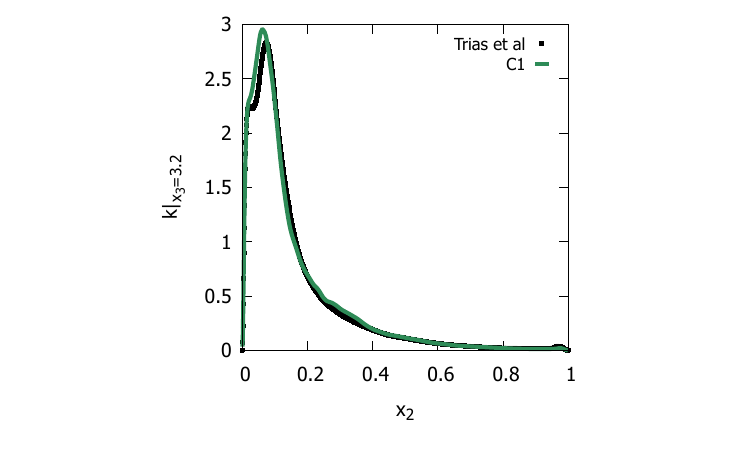}
    \captionsetup{font={footnotesize}}
    \caption{Averaged flow variables and heat transfer at $\mathrm{Ra}=10^{10}$.  }
    \label{fig:Ra10_1}
\end{figure}
To validate the code, simulations at three Rayleigh numbers $\mathrm{Ra}=6.4\times 10^6$, $2\times 10^9$ and $1\times 10^{10}$ with $64\times156\times312$, $64\times144\times318$, and $128\times138\times326$ meshes,  respectively, are compared to the DNS data. Here the resolution is identical to that of the DNS by Trias et al. \cite{TRIAS2014246,trias2007}. At $\mathrm{Ra}=10^{10}$, the length of the periodic direction is doubled here. It maybe emphasized that the current simulations A10, A11, B1 and C1 use the scalar-QR LES model described before,  and the contributions from the LES model are expected to be trivial due to the fine mesh resolution.
Figure \ref{fig:ABC} shows the horizontal temperature and velocity fields at nine vertical locations ($x_3=0.4$, $0.8$, $1.2$, $1.6$, $2.0$, $2.4$, $2.8$, $3.2$, and $3.6$) obtained from the simulations across the three Rayleigh numbers. Due to flow symmetry, only the mean temperature and velocity on the hot wall are displayed. From Figure \ref{fig:ABC}, it is evident that the mean temperature and velocity fields at each vertical location align precisely with the DNS reference data for all Rayleigh numbers. \\
Figures \ref{fig:Ra8_1}, \ref{fig:Ra9_B2_2}, and \ref{fig:Ra10_1} present key variables for the three Rayleigh numbers: the local Nusselt number distribution on the hot wall, $Nu(x_3)|_{x_2=0}$; the averaged vertical temperature profile at the mid-width cavity ($x_2=0.5$ and $x_1=0.5$); the averaged wall-shear stress scaled by $\mathrm{Ra}^{-1/4}$ on the hot wall, $\frac{\partial u_3}{\partial x_2}|_{x_2=0}$; and the turbulent kinetic energy at $x_3=3.2$. The turbulent kinetic energy in the current simulation is computed by combining the contribution from the large-eddy model with the resolved part \cite{winckelmans2002comparison}. It turns out that the contribution of the SGS model to the turbulent kinetic energy is invisible due to the fine mesh resolution. \\
%Hence, the SGS contribution to $k$ is excluded in the other test cases i.e., turbulent channel flow, flow over periodic hills and a circular cylinder and wind turbine flow in Chapter \ref{chapter:cases}.\\
The first-order flow variables and heat transfer at the three Rayleigh numbers show excellent agreement with the DNS reference data, confirming that the current numerical methods accurately predict the transition point. However, small discrepancies are observed in the second-order flow variables and temperature fields at the lowest Rayleigh number $\mathrm{Ra}=6.4\times 10^8$. Specifically, the Reynolds stress and turbulent kinetic energy are over-predicted, while temperature fluctuations are under-predicted. These differences can likely be attributed to the stretching ratio. As shown in Table \ref{tab:dhc1}, the grid size next to the wall is $(\Delta x_2)_{\text{min}}=1.63\times 10^{-3}$, which is one order of magnitude coarser than that of the reference DNS. Additionally, the order of numerical discretization schemes differs between the two methods: the DNS uses fourth-order accurate schemes, whereas the current LES employs second-order accurate schemes.
Moreover, the flow field and heat transfer are highly sensitive to initial condition perturbations. The DNS data were collected over a shorter time period compared to the simulations in this study, which may also contribute to the discrepancies. Since the simulations in this section utilize the same number of grid points as the DNS reference, we consider them an additional reference dataset for comparison in later sections.

\subsubsection{Spatial discretization at $\mathrm{Ra}=1\times 10^{10}$}
\begin{figure}[t!]
\includegraphics[trim=70 14 80 8, clip,width=.32\linewidth]{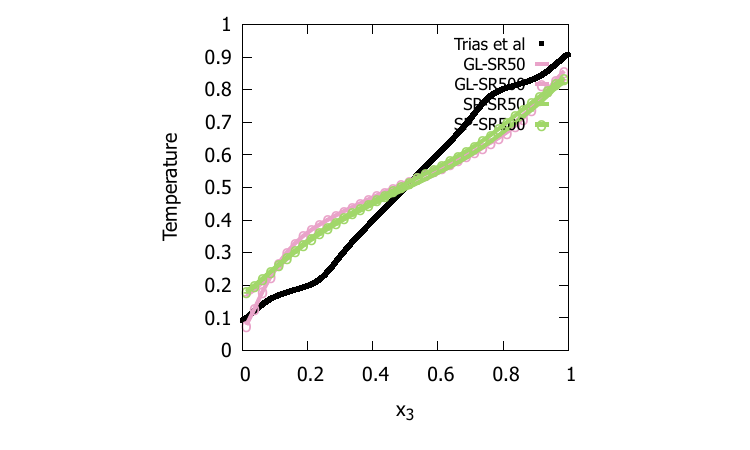}
\includegraphics[trim=70 14 80 8, clip,width=.32\linewidth]{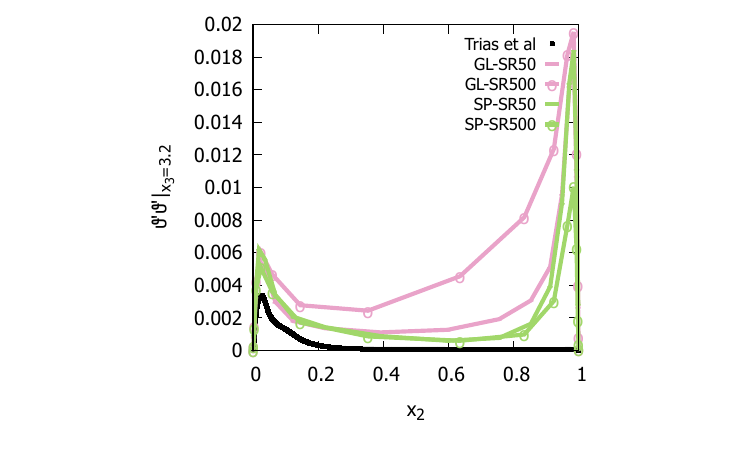}
\includegraphics[trim=70 14 80 8, clip,width=.32\linewidth]{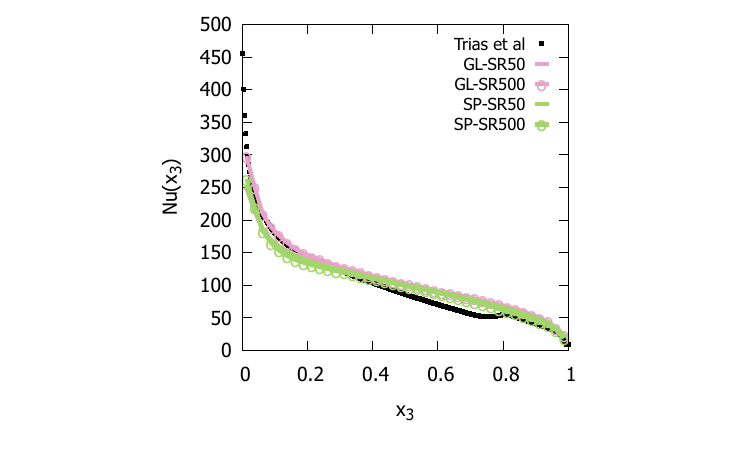}
\includegraphics[trim=70 14 80 8, clip,width=.32\linewidth]{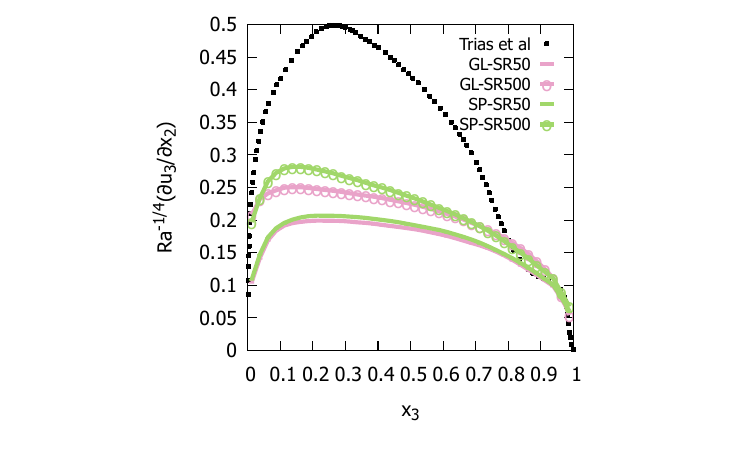}
\includegraphics[trim=70 14 85 8, clip,width=.32\linewidth]{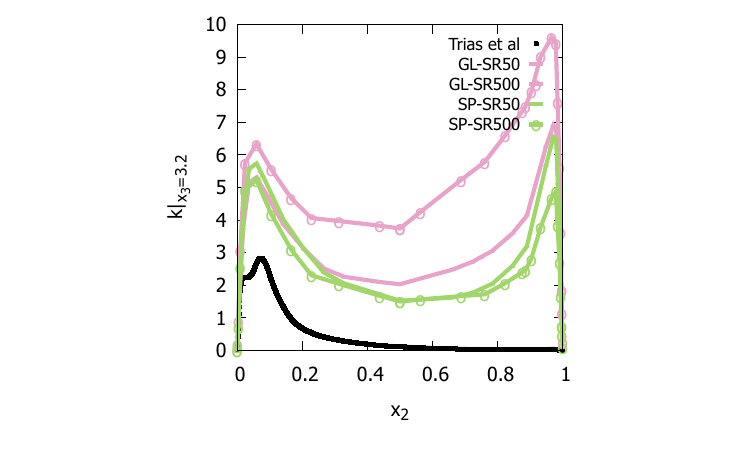}
\includegraphics[trim=70 14 85 8, clip,width=.32\linewidth]{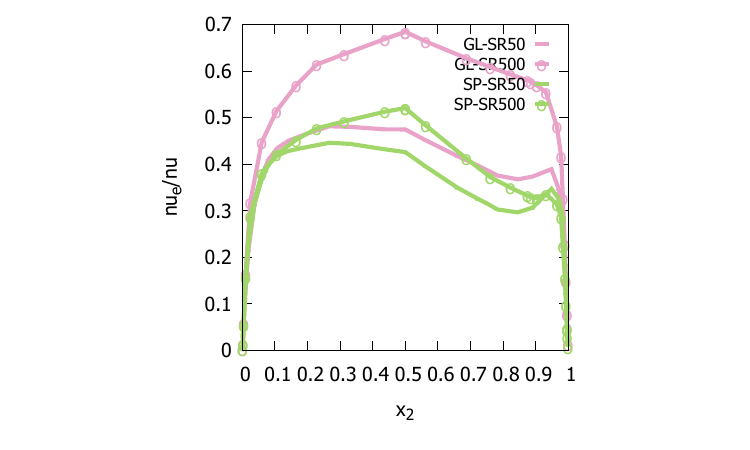}
\captionsetup{font={footnotesize}}
\caption{Flow variables and heat transfer using Gauss linear (GL) and symmetry-preserving (SP) schemes on a $8\times17\times40$ mesh with stretching ratios $SR=50$ and $500$ at $\mathrm{Ra}=1\times 10^{10}$.}
\label{fig:spCD}
\end{figure}
The results obtained using symmetry-preserving and Gauss linear \cite{OFwebsite} schemes on a $8\times17\times40$ mesh with stretching ratios of 50 and 500 are presented in Figure \ref{fig:spCD}. On meshes with a small stretching ratio ($SR=50$), the two schemes produce similar results, with symmetry-preserving discretization slightly outperforming the Gauss linear scheme.\\ 
On highly stretched meshes ($SR=500$), symmetry-preserving discretization significantly improves predictions of wall shear stress, temperature and velocity fluctuations, and turbulent kinetic energy. However, it predicts a higher averaged temperature and a lower Nusselt number along the bottom adiabatic wall. Additionally, a fixed CFL number of $CFL\leq 0.35$ yields the time step $\Delta t\approx0.01$ for the Gauss linear scheme and $\Delta t\approx0.06$ for the symmetry-preserving scheme on highly stretched meshes ($SR=500$). \\
In summary, symmetry-preserving discretization provides more accurate predictions on stretched meshes and requires less computational time. 

\subsubsection{Effects of the pressure correction methods}
\begin{figure}[t!]
    \centering
    \includegraphics[trim=141 5 140 10, clip, width=0.2\linewidth]{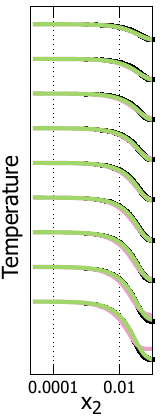}
    \includegraphics[trim=141 5 140 10, clip, width=0.2\linewidth]{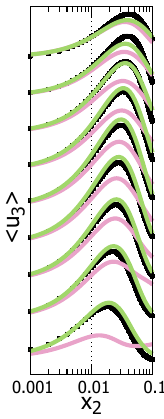}
    \includegraphics[trim=70 0 70 10, clip, width=0.42\linewidth]{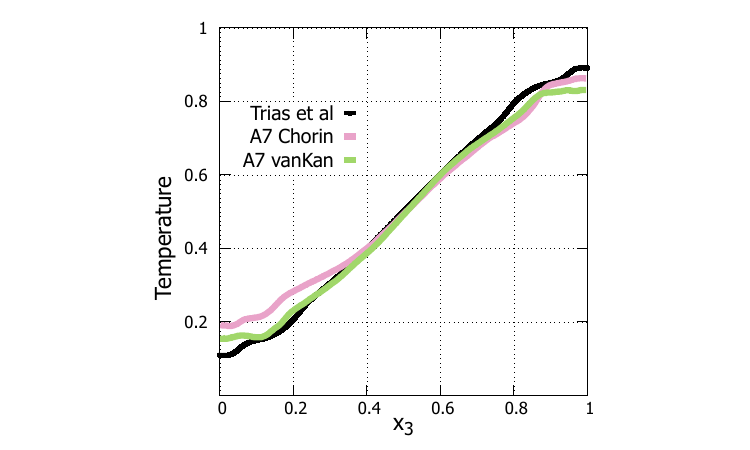} 
    \captionsetup{font={footnotesize}}
    \caption{The averaged horizontal temperature and velocity field, and vertical temperature at $\mathrm{Ra}=6.4\times 10^8$. A7vanKan and A7Chorin are the results from the cases with the second-order van Kan projection and classical Chorin projection, respectively.}
    \label{fig:pnEffect}
\end{figure}
Pressure correction methods are commonly used to reduce the computational cost of time-dependent incompressible viscous flow calculations in the velocity-pressure formulation. A key advantage of the projection method is the decoupling and sequential treatment of velocity and pressure fields. The pressure in the momentum equations acts as a projection operator, transforming an arbitrary vector field into a divergence-free vector field \cite{KIM1985308}.\\
This study compares the Chorin method \cite{chorin1968numerical} and the van Kan projection method \cite{van1986second}. The Chorin method achieves an accuracy of $O(\Delta t + \Delta x^2)$, while the van Kan method achieves $O(\Delta t^2 + \Delta x^2)$. In a simplified linearized case, the van Kan method retains the unconditional stability of the underlying symmetry-preserving scheme. Figure \ref{fig:pnEffect} illustrates the effects of these two correction methods, showing that the second-order van Kan method significantly improves the accuracy of mean temperature and velocity predictions. Notably, this method accelerates parallel computation by a factor of four in this specific case.

\subsubsection{Mesh convergence studies at  $\mathrm{Ra}=2.0\times 10^9$}
\begin{table}[b!]
    \centering
    \footnotesize
    \begin{tabular}{ccccccccc} 
    \hline
        & B1 & B2&    B3&  B4&  B5&  B6&    B7& B8 \\
    \hline
     $N_1$& 64&   64& 64& 32& 32&   16&   16&   16   \\
     $N_2$& 144&   100& 72& 72& 50& 50& 50& 32   \\
     $N_3$& 318&   256& 166& 166& 150& 150& 100&   100   \\
     $N_{tot}\times 10^6$& 2.93&   1.638& 0.765& 0.382& 0.24&  0.12& 0.08&   0.0512 \\
     SR-$x_2$&  16&  25&  25&  25&   30& 30&   30&   30  \\
     SR-$x_3$&  1.5& 1.5& 1.5& 1.5&  2& 2&  2&  2\\
     $\delta _2{_{min}} \times 10^3$&   $1.27$& $1.32 $&   $1.82$&   $1.82$& $2.27$&   $2.27$&   $2.27$&   $3.47$\\
     $\delta _2{_{max}}\times 10^2$&   $2.04$& $3.3$& $4.56$& $4.56$& $6.80$& 6.80& $6.80$&    10.4\\
     $\delta _3{_{min}} \times 10^2$&   $1.02$& $1.27$&   $1.95$& $1.95$& 1.85& 1.85& 2.77&   2.77\\
     $\delta _3{_{max}}\times 10^2$&   $1.53$&  $1.9$ &   $2.93$& $2.93$&    $3.70$& 3.70& 5.54& 5.54 \\
     $\Delta t$&0.015&0.02&0.03&0.02&0.02&0.04&0.05&0.06     \\
     $TTS$&652422 &484224 &330485&471807 &430950&238068 &216870 &157036\\
     WCT&297602  &148834&  58499&65347&44632& 26007&17337& 9853 \\  
     CPU cores& 256&  128& 128&  128& 64& 64& 64&64   \\ 
     $\frac{WCT\times CPU cores}{TTS\times N_{tot}}$&39.85&	24.01&	29.62&	46.35&	27.62&	58.26&	63.95&	78.43\\
     \hline
    \end{tabular}
    \captionsetup{font={footnotesize}}
    \caption{The cell size and computational cost in the mesh convergence studies. The computational domain has height and depth aspect ratios of $L_2/L_1=1$ and $L_2/L_3 = 4$, respectively. The unit is in meters. "SR" denotes the stretching ratio, "WCT" represents the job-wall-clock time in seconds, "TTS" is the total number of time-steps, "CPU cores" is the number of CPU cores used in parallel computing, and the last row represents the computational cost per time step and control volume in microseconds. Concerning the job-wall-clock time  in simulations B4, it is highly probable that an overhead issue occurred in this simulation.}
    \label{tab:converg1}
\end{table}
\begin{figure}[!b]
    \centering 
    \includegraphics[trim=136 0 136 2, clip, width=.21\linewidth]{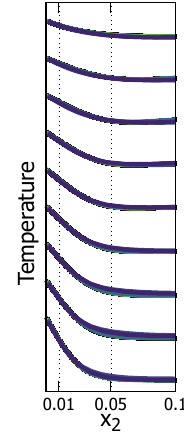}
    \includegraphics[trim=136 0 70 1, clip, width=.37\linewidth]{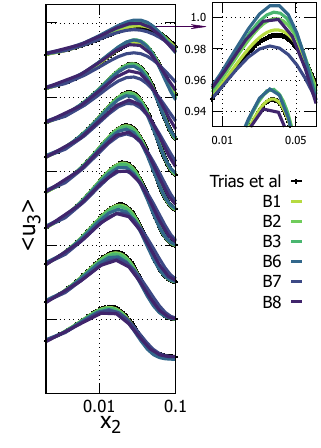}
    \captionsetup{font={footnotesize}}
    \caption{The vertical velocity (right) and averaged temperature (left) profiles obtained with various meshes at $y_3=0.4$, $0.8$, $1.2$, $1.6$, $2.0$, $2.4$, $2.8$, $3.2$, and $3.6$ at $\mathrm{Ra}=2.0\times 10^9$. Each vertical subdivision represents 0.5 units.}
    \label{fig:Ra9_BMesh_1}
\end{figure} 
\begin{figure}[!b]
    \centering
    \includegraphics[trim=68 0 72 0, clip, width=.31\linewidth]{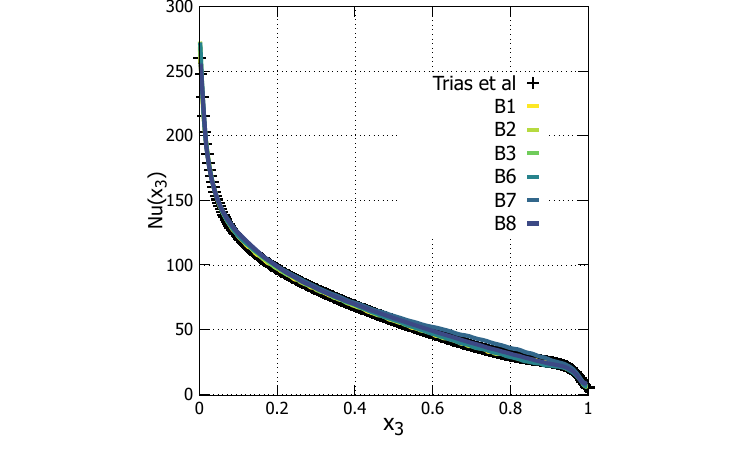}
    \includegraphics[trim=68 1 72 0, clip, width=.31\linewidth]{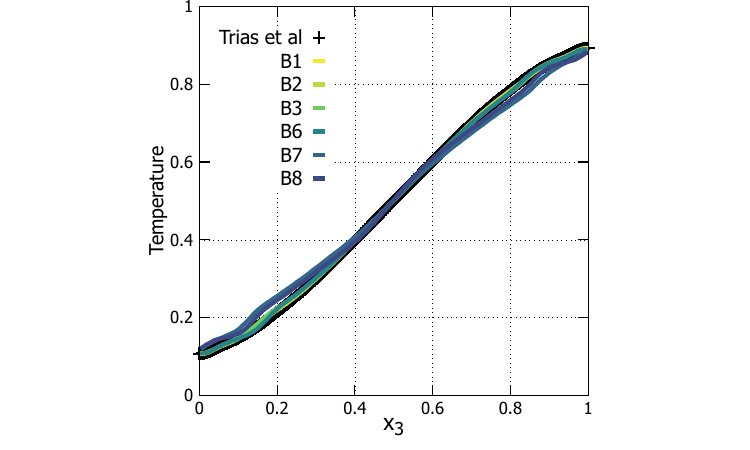}
    \includegraphics[trim=62 0 72 2, clip, width=.31\linewidth]{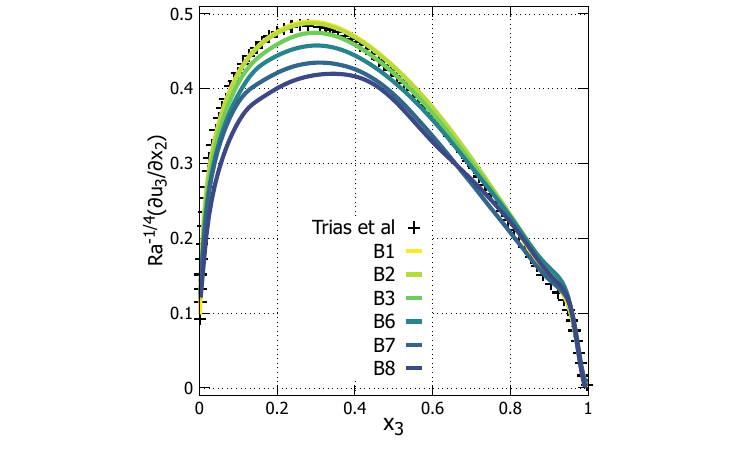} 
    \includegraphics[trim=55 5 80 2, clip, width=.315\linewidth]{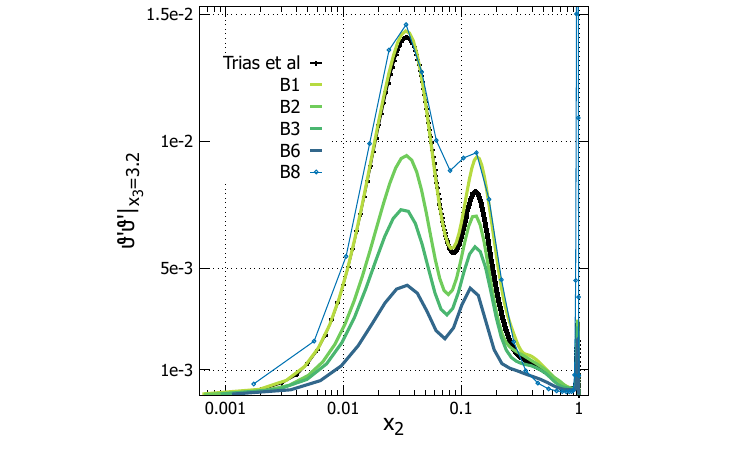}
    \includegraphics[trim=50 5 76 0, clip, width=.325\linewidth]{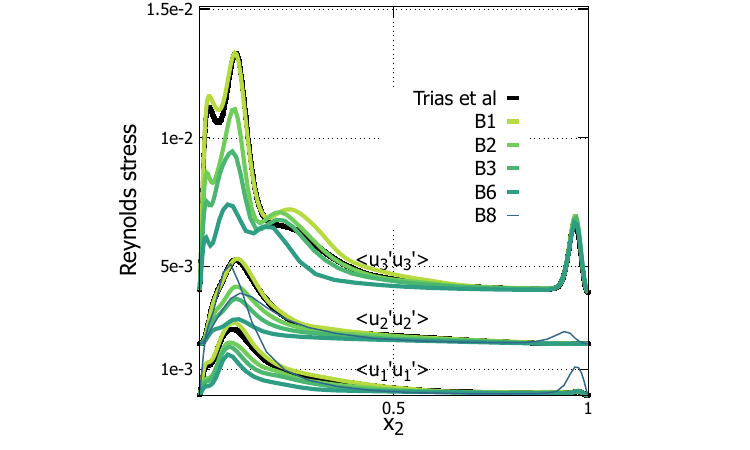} 
    \includegraphics[trim=50 0 75 0, clip, width=.31\linewidth]{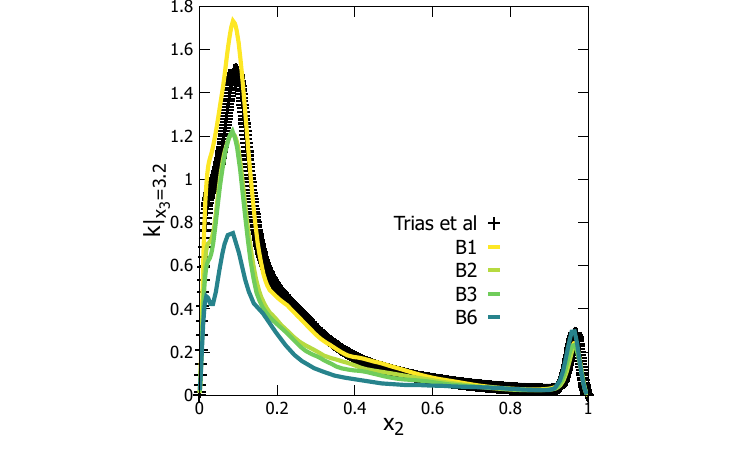}
    \captionsetup{font={footnotesize}}
    \caption{Flow variables and heat transfer obtained on different meshes at $\mathrm{Ra}=2.0\times 10^9$.}
    \label{fig:Ra9_BMesh_2}
\end{figure} 
This section addresses the mesh independence and convergence studies conducted at $\mathrm{Ra}=2.0\times 10^9$. Nine mesh configurations were tested, with statistical averages calculated over 2010 time-units after a 1340 time-units development period. Detailed information on mesh sizes and computational costs is provided in Table \ref{tab:converg1}.  The parallel computing is conducted on the Dutch National HPC system Snellius, which is operated by SURF and features AMD EPYC 7H12 (Rome) processors with 128 CPU cores per node running at 2.6 GHz, 256 GiB DRAM per node, HDR100 InfiniBand interconnect, and Lustre-based high-performance storage.\\
The results for the mean horizontal temperature and velocity fields, local Nusselt number, averaged vertical temperature, and wall-shear stress indicate that these profiles are largely insensitive to grid resolution, demonstrating the robustness of the numerical approach (LES model and discretization). However, notable discrepancies in the averaged horizontal velocity fields are observed at $x_3=3.2$ and $x_3=3.6$ in the top-left corner of the cavity, as shown in Figure \ref{fig:Ra9_BMesh_1}. Significant deviations are also found in second-order quantities, such as temperature fluctuations, velocity fluctuations, and turbulent kinetic energy. These profiles exhibit convergence to the DNS data as the grid resolution increases.\\  
For the eddy viscosity, larger values are introduced in coarse-mesh simulations to balance the production of subgrid kinetic energy. The results show that the symmetry-preserving discretization with the scalar-QR model is capable of accurately reproducing the mean flow features and heat transfer even with very coarse mesh configurations.\\
On the coarsest grid ($16\times 32\times100$), the mean temperature and velocity predictions remain accurate. However, second-order quantities, such as temperature fluctuations, are significantly under-predicted near the hot wall and over-predicted near the cold wall. This mesh configuration ($16\times 32\times100$) is therefore considered the lower resolution limit for simulations of the differentially heated cavity at $\mathrm{Ra}=2\times 10^9$.\\
Reducing the number of mesh points results in a coarser resolution in the thin vertical boundary layers near the isothermal walls. The most substantial changes occur in these boundary layers, where most heat transfer takes place, particularly in the upstream region where the flow is nearly laminar \cite{trias2007}. The scalar-QR model switches off on no-slip walls and in laminar regions, ensuring an accurate representation of near-wall behavior. Consequently, this leads to good predictions of transition points, heat transfer, and flow characteristics.  

\subsubsection{Mesh convergence study at $\mathrm{Ra}=1\times 10^{10}$}
\begin{table}[b!]
    \centering
    \small  
    \begin{tabular}{cccccccc} 
    \hline
     Case&$N_1$& $N_2$& $N_3$&   $(\Delta x_2)_{min}$& Total Time& Average Time \\ 
     \hline
    Fine&32&68&160& $1.35\times 10^{-3}$&8000&5300\\
    Medi&16&34&80& $1.26\times 10^{-3}$&10000&6700\\
    Coarse&8&17&40&$4.33\times 10^{-3}$&15000&10000\\
    \hline
    \end{tabular}
    \captionsetup{font={footnotesize}}
    \caption{The cell size and computational cost in the mesh convergence studies at $\mathrm{Ra}=1\times 10^{10}$. The computational domain has height and depth aspect ratios of $L_2/L_1=1$ and $L_2/L_3 = 4$, respectively. The mesh is stretched in $x_2$ direction. }
    \label{tab:Ra10_mesh}
\end{table}
\begin{figure}[!t]
\includegraphics[trim=70 14 80 8, clip, width=.32\linewidth]{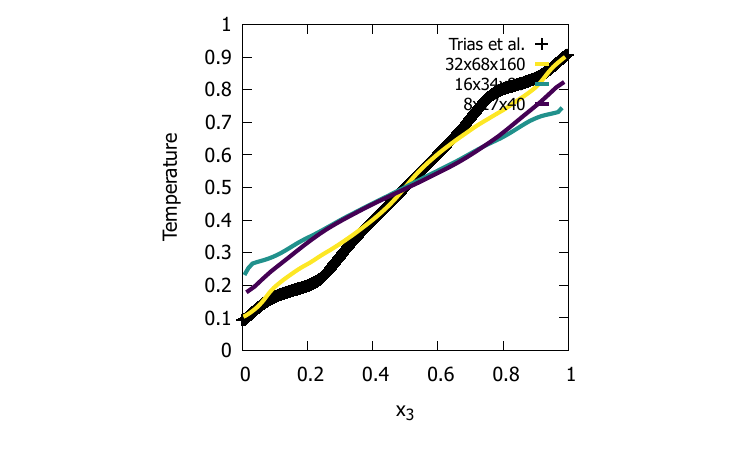}
\includegraphics[trim=70 14 80 8, clip, width=.32\linewidth]{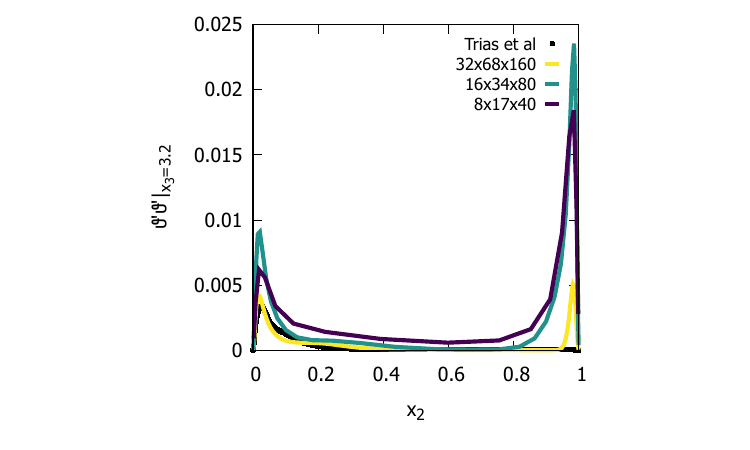}
\includegraphics[trim=70 14 80 8, clip, width=.32\linewidth]{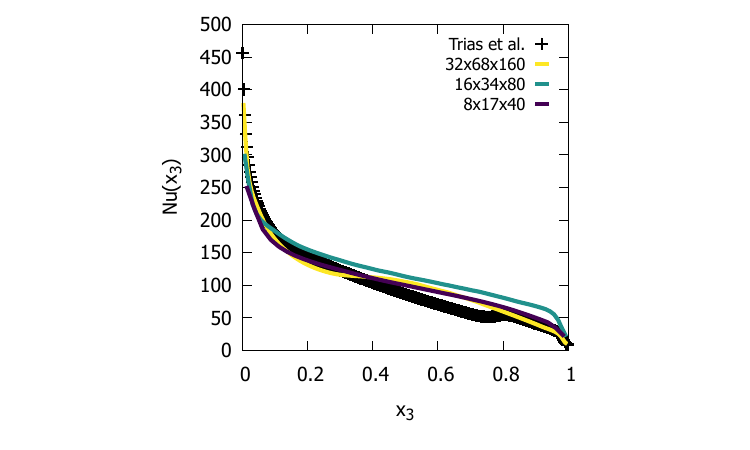}
\includegraphics[trim=70 14 80 8, clip, width=.32\linewidth]{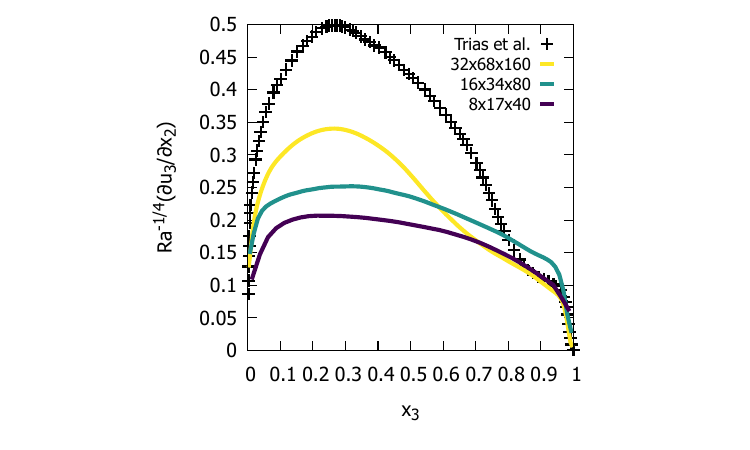}
\includegraphics[trim=70 14 80 8, clip, width=.32\linewidth]{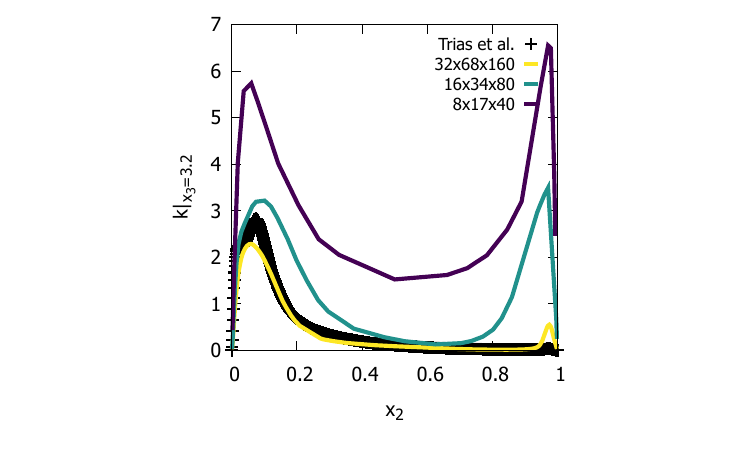}
\includegraphics[trim=70 14 80 8, clip, width=.32\linewidth]{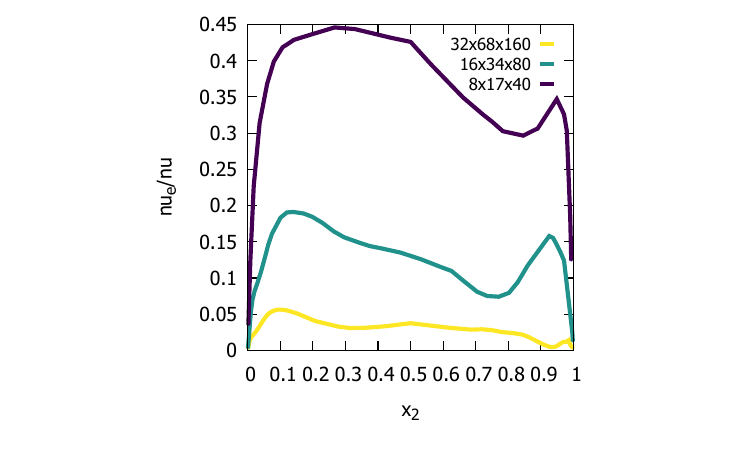}
\captionsetup{font={footnotesize}}
\caption{Flow variables and heat transfer on different meshes at $\mathrm{Ra}=1\times 10^{10}$.}
\label{fig:Ra10_mesh}
\end{figure}
A mesh convergence study was conducted using 45 different meshes ranging from five million to five thousand points, with variations in stretching ratios and the number of mesh points in different directions. Representative results for three meshes using symmetry-preserving discretization are shown in Figure \ref{fig:Ra10_mesh}, with simulation parameters detailed in Table \ref{tab:Ra10_mesh}.\\
As the mesh is refined, flow variables such as wall shear stress, velocity fluctuations, and turbulent kinetic energy monotonically converge to the DNS data. However, irregularities are observed in the heat transfer results. For the averaged vertical temperature field, the medium mesh predicts the highest temperature on the hot wall and the lowest on the cold wall, despite overlapping profiles in the cavity center ($0.8 \leq x_3 \leq 3.2$) for both medium and coarse meshes. Similarly, the medium mesh produces the highest over-prediction of peaks near the vertical isothermal walls in the temperature fluctuation field ($\vartheta'\vartheta'$), whereas predictions in the cavity center ($0.4 \leq x_3 \leq 3.2$) align accurately with DNS data.\\
For the Nusselt number, progressive convergence to the DNS data is observed along the bottom adiabatic wall. As the number of mesh points decreases, the Nusselt number gradually declines from 400 to 250, indicating reduced heat transfer. A plateau in the Nusselt number profile is observed only on the fine mesh, occurring in the cavity center around $x_3 \approx 1.5$. This plateau is shifted downward compared to the DNS plateau at $x_3 \approx 3.2$.\\
Observations of the temperature fluctuation field ($\vartheta'\vartheta'$) and turbulent kinetic energy reveal the existence of extremely thin boundary layers. On the fine mesh, turbulence and heat transfer are predominantly concentrated near the hot wall, whereas on the medium and coarse meshes, they appear along both vertical isothermal walls.\\
The eddy viscosity increases from $5\%$ of molecular viscosity on the finest mesh to $45\%$ on the coarsest mesh. Although the magnitude differs, the distribution follows a similar trend: zero at the wall, peaking at approximately $x_2=0.1$, decreasing gradually along the horizontal center, and reaching a minimum near the cold wall before rising slightly to a second peak adjacent to the cold wall.\\
In conclusion, most variables converge to the DNS data as the mesh is refined. The results on the coarse mesh ($8\times17\times40$) are sufficiently accurate, demonstrating the potential of the scalar-QR model combined with symmetry-preserving discretization in such flow scenarios.

\subsubsection{Subgrid-scale model at $\mathrm{Ra}=1\times 10^{10}$}
\begin{figure}[t!]
\includegraphics[trim=70 14 80 8, clip, width=.32\linewidth]{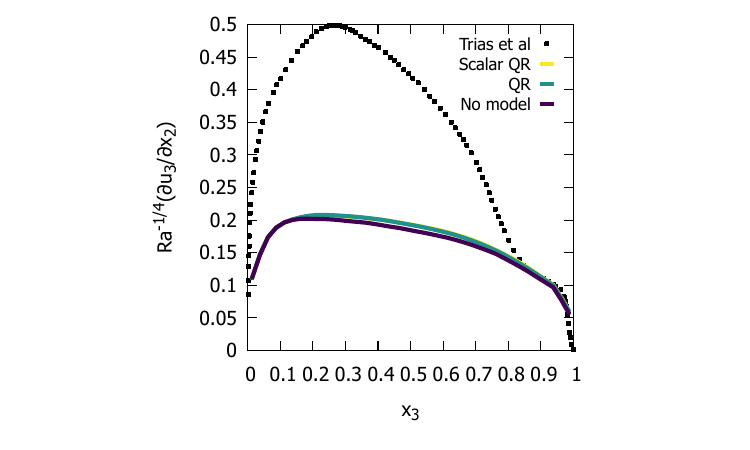}
\includegraphics[trim=70 14 80 8, clip, width=.32\linewidth]{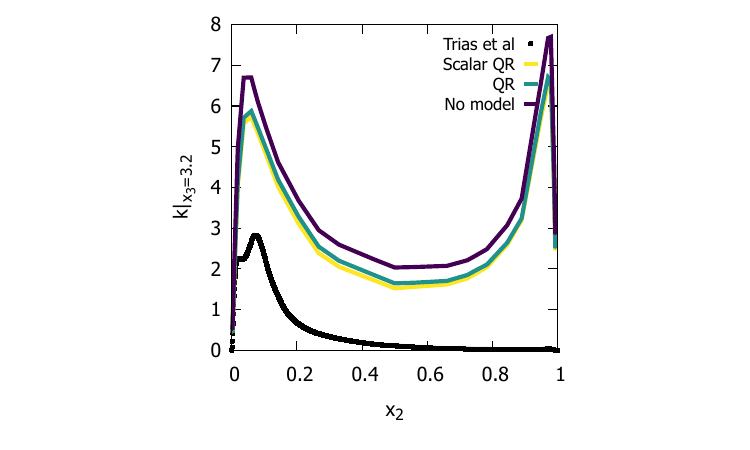}
\includegraphics[trim=70 14 80 8, clip, width=.32\linewidth]{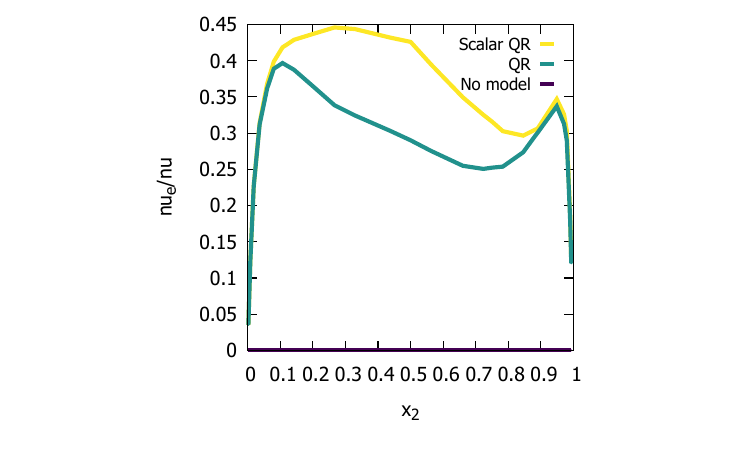}
\captionsetup{font={footnotesize}}
\caption{Flow variables and heat transfer using scalar-QR, QR and no-model using symmetry-preserving schemes on a $8\times17\times40$ mesh with stretching ratios $SR=50$.}
\label{fig:sgs}
\end{figure}
This section compares three types of LES model: the scalar-QR model, the QR model \cite{sun2024}, and the no-model approach. The QR model can be obtained by assuming the density in Eq.\eqref{eq:tempVisc} does not vary with temperature. Both symmetry-preserving and Gauss linear schemes were employed on an $8 \times 17 \times 40$ mesh with a stretching ratio of 50.\\
Using the Gauss linear scheme with $SR=500$, the scalar-QR model remains stable with a physical time step of $\Delta t = 2 \times 10^{-2}$ and provides accurate results. In contrast, the QR model becomes unstable ("blows up") with the same time step, while the no-model simulations are unstable even with a smaller time step of $\Delta t = 2.65 \times 10^{-4}$. Reducing the stretching ratio to $SR=200$ or $SR=50$, which increases the smallest mesh size, stabilizes both the scalar-QR and QR models, which then yield similar predictions for the flow variables. However, the no-model cases remain unstable even with a time step of $\Delta t = 5.5 \times 10^{-4}$.\\
With symmetry-preserving discretization, all three LES approaches are stable, even on coarse meshes with a high stretching ratio of $SR=500$. Figure \ref{fig:sgs} presents the flow variables and heat transfer results for the three LES methods, using a time step of $\Delta t = 5 \times 10^{-2}$ and $SR=50$.\\
For first-order velocity and temperature-related fields, such as the averaged temperature and the Nusselt number, the results of the scalar-QR and QR models are nearly identical and show no significant differences compared to the no-model simulation. For Reynolds stresses and turbulent kinetic energy at the most turbulence-intensive region ($x_3 = 3.2$), the scalar-QR and QR models also show minor differences, while the no-model approach significantly overpredicts values across the computational domain.\\
Regarding eddy viscosity ($\nu_e$) at $x_3 = 3.2$, the scalar-QR and QR models differ significantly in the cavity's center, where buoyancy effects play an important role. The QR model exhibits two peaks near the side walls, where shear-driven turbulence is the primary mechanism. Conversely, the scalar-QR model shows a more uniform distribution in the hot half of the cavity, where hot air rises under the combined effects of shear and buoyancy. The larger eddy viscosity introduced by the scalar-QR model in the cavity's center leads to more accurate predictions of turbulent kinetic energy. Since the scalar-QR model incorporates buoyancy effects, it dissipates more energy in the center of the cavity, improving overall accuracy.\\
In conclusion, the scalar-QR model demonstrates superior stability and accuracy when used with the Gauss linear scheme. Symmetry-preserving discretization further stabilizes simulations for the QR and no-model cases, while the scalar-QR model outperforms the original QR model in buoyancy-dominant regions.

\section{Conclusion} \label{sec:sum}
This study extends the minimum-dissipation model of LES to incorporate active and passive scalar transport mechanisms. The main findings from simulations of a differentially heated cavity at Rayleigh numbers $\mathrm{Ra}= 6.4 \times 10^8$, $2 \times 10^9$, and $10^{10}$ are summarized below
\begin{enumerate}
\item Symmetry-preserving discretization combined with the scalar-QR model accurately captures transition points, heat transfer, and flow features across a range of Rayleigh numbers.
\item At $\mathrm{Ra}= 2 \times 10^9$, the results exhibit weak dependence on grid resolution, demonstrating the robustness of the scalar-QR model and symmetry-preserving discretization for simulating thermal turbulent flows on relatively coarse meshes. 
\item At $\mathrm{Ra}= 1 \times 10^{10}$, the coarse $8 \times 17 \times 40$ mesh yields sufficiently accurate predictions and requires less computing time, indicating the potential of the scalar-QR model combined with symmetry-preserving discretization for such flow scenarios. Additionally, the scalar-QR model demonstrates greater stability with the Gauss linear scheme compared to the original QR model. Symmetry-preserving discretization stabilizes simulations for QR and no-model cases. The scalar-QR model, with its incorporation of buoyancy effects, outperforms the QR model in buoyancy-dominant regions.
\item The symmetry-preserving discretization significantly improves predictions of flow quantity and heat transfer on highly stretched meshes (SR=500).\item The second-order van Kan pressure projection method improves accuracy and achieves a fourfold speedup in parallel computation compared to the Chorin method. 
\end{enumerate}

\section*{Acknowledgements}
This work was supported by the Chinese Scholarship Council and the University of Groningen under Grant (CSC NO. 202006870021). The computing time was granted by the Dutch Research Council (NWO) under project 2022.009.
 
\section*{Data availability}
The used code is published on GitHub \cite{sunQR}. The numerical data for the channel flow can be made available upon request.

\appendix
\section{Contribution of Coriolis force to SGS model}
%%%%\section{Appendix: Contribution of Coriolis force to SGS model}
\label{appendixA}
This appendix explains why the Coriolis force term in Eq.\eqref{eq:tempVisc} vanishes. Therefore, we consider the contribution of the Coriolis force to the velocity gradient energy equation \eqref{eq:gradientEnergy} 
\begin{equation}
    \int_{\Omega_\delta }\partial _k u_i \partial_k \frac{\partial u_i}{\partial t} dx = \quad \cdot \cdot \cdot\quad + \int_{\Omega_\delta} \partial_k u_i \partial _k (f_c \epsilon  _{ij3}u_j) dx,
\end{equation}
where $\cdot \cdot \cdot$ represents the other terms on the right-hand-side of Eq.\eqref{eq:gradientEnergy}. We now evaluate the Coriolis force term
\begin{align}
    \int_{\Omega_\delta} \partial_k u_i \partial _k f_c \epsilon_{ij3}u_j dx&= f_c\int_{\Omega_\delta} \partial_k u_i \epsilon_{ij3}\partial _k  u_j dx \\
    &= f_c\int_{\Omega_\delta}( \partial_k u_1 \partial_k u_2 - \partial _k  u_2 \partial_k u_1) dx \\
    &=0.
\end{align}
The result follows from the properties of the alternating unit tensor ($\epsilon_{113} = 0$, $\epsilon_{223} = 0$, $\epsilon_{123} = 1$, and $\epsilon_{213} = -1$), which ensures that the terms cancel out.
\bibliographystyle{elsarticle-num-names} 
\bibliography{abbr_sources}
 
\end{document}